\documentclass[fleqn,usenatbib]{rasti}

\usepackage{newtxtext,newtxmath}
\usepackage[T1]{fontenc}
\DeclareRobustCommand{\VAN}[3]{#2}
\let\VANthebibliography\thebibliography
\def\thebibliography{\DeclareRobustCommand{\VAN}[3]{##3}\VANthebibliography}

\usepackage{amsmath}
\usepackage{booktabs}
\usepackage{graphicx}
\usepackage{xcolor}

\newcommand{\msun}{M_{\odot}}
\newcommand{\qf}{q_{\mathrm{F}}}

\newcommand{\qfit}{q_{\mathrm{fit}}}

\newcommand{\snrarc}{\mathrm{SNR}_{\rm arc}}
\newcommand{\gabs}{G_{\rm abs}}
\newcommand{\grel}{G_{\rm rel}}

\title[HWO strong-lensing subhalo sensitivity]{Point spread function requirements for
dark matter subhalo detection with the Habitable Worlds Observatory}

\author[G.\ N.\ Vassilakis et al.]{\fontsize{11.6}{15}\selectfont Georgios N.\ Vassilakis$^{1,2,3,4}$\thanks{E-mail: gv321@cam.ac.uk (GNV)},
Susan F.\ Redmond$^{4}$,
Jason D.\ Rhodes$^{4}$,
Richard J.\ Massey$^{5}$,
and Qiuhan He$^{6,5}$
\\
$^{1}$Institute of Astronomy, University of Cambridge, Cambridge, United Kingdom\\
$^{2}$Department of Applied Mathematics and Theoretical Physics, University of Cambridge, Cambridge, United Kingdom\\
$^{3}$Kavli Institute for Cosmology, University of Cambridge, Cambridge, United Kingdom\\
$^{4}$Jet Propulsion Laboratory, California Institute of Technology, Pasadena, CA, United States\\
$^{5}$Institute for Computational Cosmology, Department of Physics, Durham University, Durham, United Kingdom\\
$^{6}$Kapteyn Astronomical Institute, University of Groningen, Groningen, the Netherlands
}

\date{Submitted to RAS Techniques and Instruments (RASTI) for the Habitable Worlds Observatory special issue.}
\pubyear{2026}
\renewcommand{\today}{1 October 2026}

\begin{document}
\label{firstpage}
\pagerange{\pageref{firstpage}--\pageref{lastpage}}
\maketitle

\begin{abstract}
Different theories of dark matter predict different populations and morphologies of galactic subhaloes below about $10^8\msun$. The planned angular resolution and optical stability of the Habitable Worlds Observatory (HWO) should enable it to detect subtle subhalo perturbations in strongly lensed galaxy images. We simulate 2000-second exposures with its High Resolution Imager (HRI) under the fiducial Exploratory Analytic Case 1 (EAC1) design, assuming known source structure and omitting lens-galaxy light. Pre-selecting the top 50 of 965 lenses (5.2\%) using only arc signal-to-noise and relative gradient power lowers the median mass threshold for 10\% aperture coverage roughly tenfold relative to the full ensemble. The top 12 (1.2\%) reach a median of $6.6\times10^7\msun$ for the same coverage. At their most sensitive positions, the top 50 and top 12 reach median thresholds of $3\times10^7\msun$ and $1.5\times10^7\msun$, respectively. Non-linear fits support the Fisher forecasts at the tested masses and positions. For these 12 lenses, increasing the physical wavefront error from 35 to 65~nm RMS raises the median thresholds by only 0.1~dex when the point spread function (PSF) is modelled correctly. Deliberately mismodelling the 35-nm reference PSF removes detectable positions, although false detections remain rare. Under our illustrative retention criteria, loss of sensitive area sets a median tolerated wavefront mismatch of 5~nm RMS (range 2--10~nm). Selective HWO follow-up of lenses discovered by Euclid or Roman could therefore transform our understanding of the nature of dark matter, provided that the PSF is characterized accurately enough to preserve this sensitivity.
\end{abstract}


\section{Introduction}
\label{sec:intro}

Various theories of dark matter predict vastly different morphologies and populations of dark matter subhaloes. In a nearly homogeneous universe with small overdensities, an overdense region can attract surrounding matter. Cold dark matter (CDM) predicts dark matter particles with small primordial random velocities, where negligible free streaming on galactic scales is expected. This leads to CDM collapsing into these overdense regions, where some merge into the larger galaxy-sized haloes and others remain as smaller subhaloes \citep{bullock2017_small_scale_challenges}. Warm dark matter (WDM) predicts dark matter particles with larger primordial random velocities compared to CDM, leading to larger free streaming \citep{bode2001_wdm}. This leads to the suppression of small-scale primordial overdensities as particles stream out of the smallest overdense regions, reducing the density contrast before gravity can amplify these regions into haloes. Small-scale fluctuations are smoothed, while larger-scale overdensities cannot be sufficiently suppressed in time, and instead form larger galaxy-sized haloes. As a result, WDM predicts a much smaller number of low-mass subhaloes compared to CDM \citep{lovell2014_wdm_haloes}. Therefore, measuring subhalo populations, especially at low masses, can be a powerful probe into the theoretical nature of dark matter \citep{vegetti2014_inference}.

Gravitational lensing has proven to be a powerful tool for detecting these dark matter subhaloes \citep{vegetti2010_dark_substructure}. The 'strong' gravitational lensing regime is characterized by the extreme distortions of background sources that are closely aligned with the lensing object in the line of sight. These sources can form extended arcs of light around the lensing object when the background source is extended, or multiple imaging when the background source is compact, such as a quasar. We can directly model the paths the light rays from these sources must travel around the lens galaxy using general relativity \citep{bartelmann2010_lensing}. But when dark matter subhaloes are present in the lens galaxy, these subhaloes can act as additional lenses of the background source light, creating localized distortions in the spatial brightness pattern. We can numerically model these distortions and thus can detect the presence of these low-mass perturbers to attempt to distinguish between competing dark matter models \citep{vegetti2009_bayesian}.

However, detection of low-mass subhaloes in the regime where these theories meaningfully diverge is challenging. The lensing system has many parameters to consider, and all must be inferred jointly to search for the substructure \citep{vegetti2009_bayesian}. These components can explain part of the suspected subhalo perturbation signal: adjusting the macro lens model changes the lensing geometry, and adjusting the source model changes the brightness structure that is being mapped \citep{despali2022_sensitivity}. In a similar vein, overly restrictive model assumptions can generate confident, but incorrect, subhalo evidence \citep{he2022_cosmo_subhalo_detection}. Therefore, trustworthy analysis requires physically justified model flexibility, even if this reduces the detection significance. Fine structure that carries information about low-mass perturbers can be further degraded by finite angular resolution \citep{despali2022_sensitivity}, while fine detector sampling is also needed to preserve that spatial information delivered by the optical system. Point spread function (PSF) variations can also alter the apparent substructure signal \citep{wedig2025_roman_lenses}. Consequently, these challenges motivate observations that combine high angular resolution and sufficient signal-to-noise with a stable, accurately characterized PSF.

The upcoming Habitable Worlds Observatory (HWO) may be an ideal instrument to fill many of these requirements. HWO is a NASA flagship-level mission concept designed in response to the Astro2020 decadal survey \citep{nasem2023_astro2020}. Its primary goal is to directly image nearby Earth-like exoplanets, enabled by extremely high-contrast coronagraphy and extremely stable instrument optics \citep{liu2026_hwo_eacs}. Although its name highlights the search for habitable worlds, HWO is also intended to support a broad astrophysics programme \citep{dressing2026_hwo_discovery}. Candidate instruments for this broader programme include a High Resolution Imager (HRI) and an ultraviolet multi-object spectrograph (UV-MOS), enabling investigations from planetary to cosmological scales \citep{liu2026_hwo_eacs}. With the space-based, ultra-stable optics required for the coronagraphy alongside the HRI, HWO is well suited to carry out a low-mass subhalo detection programme to distinguish between various theories of dark matter \citep{he2026_hwo25_strong_lensing}.

Even with HWO's proposed capabilities, the sensitivity to low-mass subhaloes will vary between different lens systems. \citet{despali2022_sensitivity} have shown that even with the same observational setup, maximum sensitivity can depend dramatically on the source structure and lens geometry. \citet{wedig2025_roman_lenses} have forecasted that the Nancy Grace Roman Space Telescope's High Latitude Wide Area Survey (HLWAS) alone will discover approximately 160,000 lensing systems, which would motivate a future HWO strong lensing programme to perform follow-up observations over the systems with the highest expected subhalo information content \citep{oriordan2023_euclid_sensitivity}. PSF quality will be a critical factor in the success of detecting these low-mass subhaloes with statistical significance. Two distinct optical questions for forecasting such a programme arise: how does the sensitivity change when the physical PSF degrades but remains correctly modelled (PSF quality), and how does it change when the PSF is mismodelled (PSF knowledge error)?

In this work, we quantify how mass reach varies across lensing systems and how PSF quality and modelling accuracy affect subhalo detection sensitivity for HWO's Exploratory Analytic Case 1 (EAC1) architecture \citep{liu2026_hwo_eacs}. To simulate and forecast HWO's subhalo detection sensitivity case, we use the HWO Strong Lensing Analysis Pipeline for Subhaloes (HWO-SLAPS) \citep{vassilakis2026_hwo_slaps}. This pipeline, which is open-source on GitHub\footnote{\href{https://github.com/nasa-jpl/hwo-slaps}{https://github.com/nasa-jpl/hwo-slaps}}, simulates the subhalo detection sensitivity case end-to-end. HWO-SLAPS builds a galaxy-galaxy lensing system, injects or omits a subhalo into the lens galaxy, then convolves that scene with selected HWO-like PSFs and adds detector and measurement noise to create a mock observation. These mock observations are then analysed through two complementary approaches: fast Fisher/Asimov likelihood-ratio forecasts \citep{cowan2011_asymptotic} and targeted non-linear model comparisons, which yield likelihood ratios \citep{nightingale2021_pyautolens,lange2023_nautilus}. In this study, we generate 1000 lensing systems spanning a range of source and lens properties and compute Fisher sensitivity maps for the 965 systems that pass our observability cuts. We test these forecasts with targeted non-linear fits to the 50 highest-ranked systems and a pseudorandom validation sample of 100 systems, including no-subhalo controls to assess false-positive detections. We also explore the effect of the PSF on these detection metrics through two separate arms: varying physical PSF quality with correct modelling and introducing controlled PSF knowledge errors. Under declared observing conditions, these foreground-free simulations provide a controlled baseline for exploring the feasibility of a potential HWO low-mass subhalo detection programme.

The paper proceeds as follows: Section~\ref{sec:simulation} describes the simulation suite and lens population. Section~\ref{sec:metrics} defines the detection metrics and validation protocol. Section~\ref{sec:sensitivity} presents the mass-reach and target-selection results alongside the non-linear validation set. Section~\ref{sec:psf} examines the effects of PSF quality and knowledge errors on subhalo detection sensitivity. We discuss the implications and limitations in Section~\ref{sec:discussion} and summarize our conclusions in Section~\ref{sec:conclusion}.

\section{Simulation suite and lens population}
\label{sec:simulation}

In this section, we describe the simulation suite and lens population, which spans a range of lens and source properties under a fiducial EAC1/HRI observation configuration. We then describe the simulated physical PSFs and how we introduce controlled errors in the PSF model to later explore the effects of PSF quality and knowledge errors on subhalo detection sensitivity (Section \ref{sec:psf}).

\subsection{Instrument and observing configuration}
\label{subsec:observing}

NASA has been developing candidate designs, known as Exploratory Analytic Cases (EACs) to evaluate the feasibility of various HWO mission concept designs. We use EAC1 as our fiducial telescope design as described in \citet{liu2026_hwo_eacs}, along with the HRI instrument inputs from version 0.1.9 of the HWO Science Engineering Interface (SEI)\footnote{\href{https://github.com/HWO-GOMAP-Working-Groups/Sci-Eng-Interface/blob/1af10b18c4988340dc007aeb3442877ae5b477f6/hwo_sci_eng/obs_config/HRI/HRI.yaml}{SEI HRI configuration file} (commit \texttt{1af10b1}).}. These inputs describe a provisional design, and the configurability of the HWO-SLAPS pipeline allows this analysis to be repeated with revised telescope and instrument designs in future studies. The key simulation parameters describing the assumed HWO EAC1/HRI instrument configuration are listed in Table~\ref{tab:observing_configuration}. Wavefront errors are expressed as root-mean-square (RMS) values.

\begin{table}
\centering
\caption{Simulation parameters for the foreground-free EAC1/HRI ultraviolet--visible (UVIS) mock observations. Pupil geometry, pixel scale, per-read noise and dark current follow SEI v0.1.9, while other values are adopted or derived.}
\label{tab:observing_configuration}
\small
\begin{tabular}{@{}p{0.49\columnwidth}p{0.46\columnwidth}@{}}
\toprule
Parameter & Adopted value \\
\midrule
\multicolumn{2}{@{}l}{\textbf{Telescope and pupil}} \\
\addlinespace[0.4ex]
Primary mirror & 19 hexagonal segments \\
Inscribed pupil diameter & $5.977\,\mathrm{m}$ \\
Circumscribed pupil diameter & $7.226\,\mathrm{m}$ \\
Collecting area & $33.606\,\mathrm{m}^{2}$ \\
\addlinespace[1.0ex]
\multicolumn{2}{@{}l}{\textbf{Observing configuration}} \\
\addlinespace[0.4ex]
Photometric band & $450$--$550\,\mathrm{nm}$ \\
Exposure time & $2000\,\mathrm{s}$ \\
Pixel scale & $0.00716\,\mathrm{arcsec\,pixel}^{-1}$ \\
End-to-end throughput & $0.21$ \\
\addlinespace[0.4ex]
\multicolumn{2}{@{}l}{\textbf{Point spread function}} \\
\addlinespace[0.4ex]
PSF-rendering wavelength & $500\,\mathrm{nm}$ \\
Reference wavefront error & $35\,\mathrm{nm}$ RMS \\
\addlinespace[0.4ex]
\multicolumn{2}{@{}l}{\textbf{Detector and background}} \\
\addlinespace[0.4ex]
Read noise per frame & $0.2\,\mathrm{e}^{-}$ \\
Number of reads & $2$ \\
Combined read noise & $0.283\,\mathrm{e}^{-}$ \\
Dark-current rate & $2.0\times10^{-3}\,\mathrm{e}^{-}\,\mathrm{pixel}^{-1}\,\mathrm{s}^{-1}$ \\
Sky surface brightness (AB) & $23.0\,\mathrm{mag\,arcsec}^{-2}$ \\
Sky-background rate & $2.5103\times10^{-3}\,\mathrm{e}^{-}\,\mathrm{pixel}^{-1}\,\mathrm{s}^{-1}$ \\
\bottomrule
\end{tabular}
\end{table}

All lens systems share a common exposure time and pixel scale, allowing their sensitivity to be compared under the same observing conditions. This defines a fiducial observation for the study, rather than a proposed HWO observing schedule. We calculate photon rates over the adopted photometric band, assuming a constant flux density per unit frequency within that band. The PSF is instead rendered at the central wavelength of 500~nm, so the image response is monochromatic, even though the count rates include the full band. We describe the PSF construction in Section~\ref{subsec:psf_model}.

The source photometry, collecting area, and effective throughput determine the detected source count rates, which are then scaled by the exposure time. We adopt an effective throughput of 0.21, including detector quantum efficiency, consistent with the SEI EAC1/HRI Al+XeLiF optical chain. This retains the LUVOIR HDI reference value \citep{luvoir2019_final_report} and corresponds to an assumed filter transmission of approximately 0.83 in the SEI chain. We apply this throughput once when deriving the source and sky rates. The source normalization is described in Section~\ref{subsec:sources}. Our noise model includes Poisson fluctuations in the source, uniform sky background, and dark-current counts, together with Gaussian detector read noise. We assume two independent reads, so their variances add to give the combined read noise in Table~\ref{tab:observing_configuration}. The total variance in each pixel is the sum of the expected source, sky, and dark-current electron counts and the read-noise variance.

In these lensing simulations, we omit the light from the lens galaxy while retaining its gravitational effect. This includes omitting its photon noise and any residuals from light subtraction. This provides a controlled baseline for comparing subhalo sensitivity across lens systems and PSF conditions. The following subsection describes the mass models used to generate these lensing systems.

\subsection{Lens and subhalo models}
\label{subsec:lens_subhalo}

Within the HWO-SLAPS pipeline, we use PyAutoLens \citep{nightingale2021_pyautolens} to model and simulate all aspects of the lensing system, including the source, lens, and subhalo. We model the lens galaxy as having a large-scale mass distribution, denoted the macro lens, which is chiefly responsible for the strong lensing effect, while a smaller-scale mass profile, denoted the subhalo, lives within the macro lens and is optionally added to the simulation. We represent the macro lens as a singular isothermal ellipsoid (SIE) mass distribution \citep{kormann1994_sie}, whose dimensionless projected surface mass density is given by

\begin{equation}
\begin{aligned}
\kappa(\xi) &= \frac{\Sigma(\xi)}{\Sigma_{\rm crit}}
              = \frac{\theta_{\rm E,eff}}{2\xi},
\qquad \Sigma_{\rm crit} = \frac{c^2}{4\pi G}\frac{D_{\rm s}}{D_{\rm l}D_{\rm ls}}, \\
\xi^2 &= q_{\rm lens}\,x'^2 + \frac{y'^2}{q_{\rm lens}},
\end{aligned}
\label{eq:sie_profile}
\end{equation}
where $\kappa$ is the convergence, $\Sigma$ is the projected surface mass density, and $\Sigma_{\rm crit}$ is the critical surface density for the lens--source system. Here $D_{\rm s}$, $D_{\rm l}$, and $D_{\rm ls}$ are the observer--source, observer--lens, and lens--source angular-diameter distances, respectively. The constants $c$ and $G$ are the speed of light and Newton's gravitational constant, respectively. The angular coordinates $x'$ and $y'$ are centred on the lens and aligned with its projected major and minor axes, respectively, with $q_{\rm lens}$ denoting the minor-to-major axis ratio and $\xi$ the corresponding elliptical radius. We define $\theta_{\rm E,eff}=\sqrt{A_{\rm crit}/\pi}$ as the area-equivalent Einstein radius, where $A_{\rm crit}$ is the angular area enclosed by the tangential critical curve.

The adopted cosmology is the spatially flat $\Lambda$CDM model implemented as Planck15 within PyAutoLens, with $H_0=67.74\,\mathrm{km\,s^{-1}\,Mpc^{-1}}$ and $\Omega_{\rm m}=0.3075$ \citep{planck2015_cosmological_parameters}. The SIE represents the combined smooth mass of the lens galaxy, including both baryonic and dark matter, without an additional external-shear component. We keep the SIE model family fixed, while allowing its centre, mass normalization, and ellipticity to adjust in the non-linear search, as described in Section~\ref{sec:metrics}.

The subhalo follows a spherical Navarro--Frenk--White (NFW) profile \citep{navarro1997_nfw} and is placed at the same redshift as the lens galaxy. Its three-dimensional density is given by
\begin{equation}
\rho(r)=\frac{\rho_s}{(r/r_s)(1+r/r_s)^2},
\label{eq:nfw_profile}
\end{equation}
where $r$ is the physical radius from the subhalo centre, $r_s$ is the scale radius, and $\rho_s$ sets the density normalization. The density approaches $\rho\propto r^{-1}$ at small radii and $\rho\propto r^{-3}$ at large radii, with a logarithmic slope of $-2$ at $r=r_s$. We define the subhalo mass ($M_{200}$) and concentration ($c_{200}$) as

\begin{equation}
\begin{aligned}
M_{200} &= \frac{4\pi}{3}\,200\,\rho_{\rm crit}(z_l)\,r_{200}^3, \\
\rho_{\rm crit}(z_l) &= \frac{3H(z_l)^2}{8\pi G}, \\
c_{200} &= \frac{r_{200}}{r_s},
\end{aligned}
\label{eq:halo_mass_concentration}
\end{equation}
where $r_{200}$ is the physical radius enclosing a mean density of $200\rho_{\rm crit}(z_l)$, and $H(z_l)$ is the Hubble parameter at the lens redshift. Here $\rho_{\rm crit}$ is the cosmological critical volume density, distinct from the lensing critical surface density $\Sigma_{\rm crit}$. The concentration $c_{200}$ is the ratio of $r_{200}$ to the scale radius $r_s$. At fixed $M_{200}$ and redshift, a higher concentration $c_{200}$ places more of the mass within the smaller projected radii, strengthening the lensing perturbation which can increase the detection significance \citep{amorisco2022_halo_concentration}.

We set $c_{200}$ using the concentration mass relation defined in equation~(7) of \citet{moline2017_subhalo_concentrations}, which depends on $M_{200}$ and $x_{\rm sub}$, the three-dimensional distance from the host centre in units of the host virial radius. This relation assigns higher concentrations to subhaloes that are closer to the centre of the host halo. We set $x_{\rm sub}=1$ independent of the subhalo's projected position, which is a conservative choice compared to higher concentrations predicted for subhaloes that are embedded deeper within the host. The relation was calibrated at $z=0$ for Milky Way-mass hosts, and we apply it without any redshift-evolution term or intrinsic scatter. The lens redshift therefore only enters through $\rho_{\rm crit}(z_l)$ in equation~\eqref{eq:halo_mass_concentration} and through the lensing distances in $\Sigma_{\rm crit}$ in equation~\eqref{eq:sie_profile}.

Throughout this study, subhalo mass refers to $M_{200}$, the mass enclosed within $r_{200}$. The spherical NFW profiles remain untruncated, as in \citet{nightingale2023_subhalos}. The subhalo mass and projected position are varied systematically to map detection sensitivity across the lensing systems described in Section \ref{subsec:selection}. This suite of simulations considers the effects of placing a single subhalo at the lens redshift, with no additional population of haloes along the line of sight. 

\subsection{Source models}
\label{subsec:sources}

\begin{figure*}
    \centering
    \includegraphics[width=\textwidth,alt={Five columns show clumpy, barred-spiral, irregular, compact, and smooth-disc source galaxies above their corresponding lensed images. Lensing stretches their brightness structure into curved arcs around a dark centre, preserving differences in knots, spiral features, and smoothness.}]{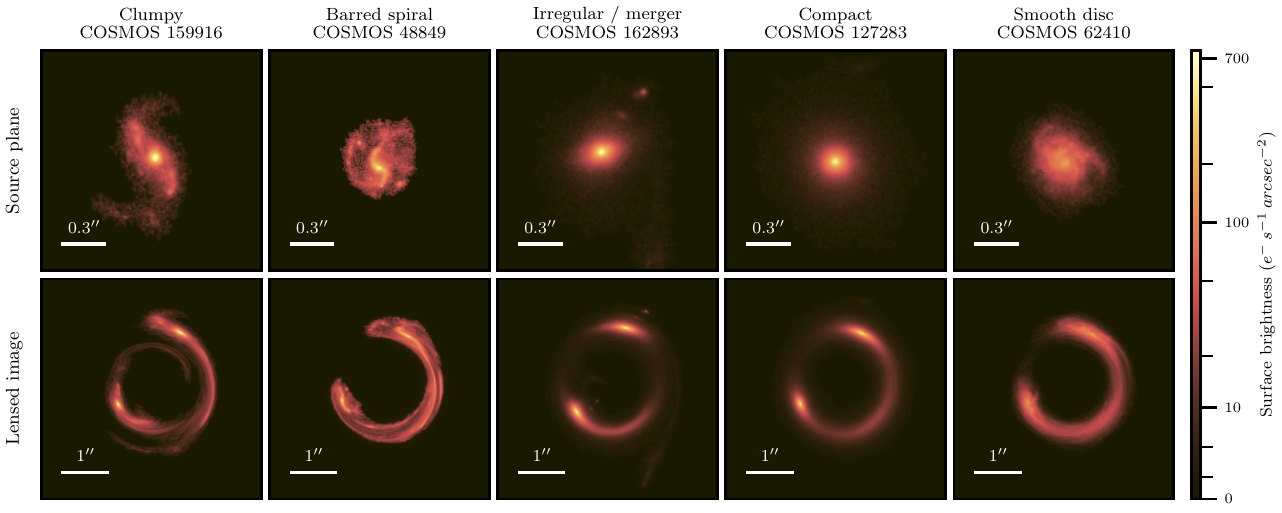}
    \caption{The five COSMOS source templates (top) and their lensed images (bottom), shown using a common configuration constructed from the marginal medians of the initial 1000-system pool. The lensed images include the reference PSF at 500~nm with 35~nm RMS wavefront error. No subhalo, lens-galaxy light, or additional noise realization is included. From left to right, the templates are COSMOS 159916, 48849, 162893, 127283, and 62410.}
    \label{fig:source_lensing_gallery}
    \end{figure*}

Different source morphologies will lead to different gradient structures in the lensed image, and this can affect the detection significance of the subhalo \citep{despali2022_sensitivity}. Motivated by this, we use five source templates spanning a clumpy galaxy, a barred spiral, an irregular galaxy, a compact galaxy, and a smooth disc. These templates are based on F814W images taken with the Advanced Camera for Surveys (ACS) on the Hubble Space Telescope (HST) for the COSMOS survey \citep{koekemoer2007_cosmos_acs}, obtained from the real-galaxy sample distributed for GalSim \citep{rowe2015_galsim}. The five source templates are shown in the top row of Fig.~\ref{fig:source_lensing_gallery}.

Each image is background-subtracted, masked to isolate the galaxy, and centred on its light centroid, with negative pixel values set to zero. The angular scale is then adjusted so that a circular aperture of radius 0.11~arcsec contains half of the retained light. This provides a common reference size before the source sizes, orientations, and positions are assigned as described in Section~\ref{subsec:selection}. During ray tracing, bilinear interpolation gives the source surface brightness at each mapped source-plane position. The templates are used without deconvolving the original HST PSF and retain residual observational noise.

We anchor the source brightness to the average unlensed F814W magnitude of 24.3 reported in the Sloan Lens ACS (SLACS) survey \citep{newton2011_slacs}. An adopted colour of $m_{\rm HRI-V}-m_{\rm F814W}=0.545$ gives a reference HRI-V magnitude of 24.845. Each template is initially normalized to the corresponding unlensed detected rate of $8.95\,\mathrm{e}^{-}\,\mathrm{s}^{-1}$ under the observing conditions in Section~\ref{subsec:observing}. The magnitudes and sizes of individual sources are drawn as described in Section~\ref{subsec:selection}. When a source is resized, its surface-brightness normalization is adjusted to preserve the total flux at its assigned magnitude. Lensing magnification is generated by ray tracing and is not applied during source normalization. The five templates thus provide a controlled range of source structures for the mass reach analysis, without modelling the full source population or changes in morphology with wavelength or redshift.

\subsection{Lens population and target selection}
\label{subsec:selection}

We generate an initial pool of 1000 lensing systems using the sampling prescriptions listed in Appendix~\ref{app:population_sampling} (Table~\ref{tab:population_parameters}). We balance the allocation of systems by source morphology, assigning 200 systems per template. The pool allows the exploration of a controlled range of lens geometries and source properties to compare mass reach and target selection.  

The lens and source redshift distributions are centred at $z_{\rm l}=0.22$ and $z_{\rm s}=0.65$, respectively, motivated by the typical SLACS lens and source redshifts \citep{newton2011_slacs}. The lens velocity dispersion and lens-source distances determine the Einstein radius of the lens. Source placement is specified by their angular offset from the lens centre, which we express as a fraction of this radius. We draw this fraction from a log-uniform distribution, emphasizing close alignments within the strong lensing regime. Source magnitudes are drawn from a normal distribution centred on the SLACS-based reference magnitude of 24.845 described in Section~\ref{subsec:sources}. The full sampling distributions and their bounds are provided in Appendix~\ref{app:population_sampling}.

Next, we identify targets from the initial 1000-system pool that are likely to have the highest low-mass subhalo sensitivity, using only their images before any sensitivity is measured. This represents a follow-up style programme in which HWO observes the highest-ranked lensing systems from a population of  hundreds of thousands of systems, as expected from Euclid and Roman's lensing surveys \citep{oriordan2023_euclid_sensitivity, wedig2025_roman_lenses}. Bright arcs provide more signal and brightness gradients can make small deflections more visible, so we construct a selection score based on these properties. Each system is rendered without a subhalo, using the observing conditions described in Section~\ref{subsec:observing}. From this expected image, we calculate the integrated arc signal-to-noise ratio $\snrarc$, noise-weighted gradient power $\gabs$, and dimensionless relative gradient power $\grel$ within a circular aperture of radius $2\theta_{\rm E,eff}$ centred on the lens:

\begin{equation}
\begin{aligned}
\snrarc^2 &= \sum_i \frac{s_i^2}{\sigma_i^2}, \\
\gabs &= \sum_i \frac{|\nabla s_i|^2}{\sigma_i^2}, \\
\grel &= \frac{\theta_{\rm res}^2 \gabs}{\snrarc^2},
\end{aligned}
\label{eq:selection_quantities}
\end{equation}

where $s_i$ is the expected source electron count in pixel $i$ and $\sigma_i^2$ is its variance under the noise model in Section~\ref{subsec:observing}. The gradient is taken with respect to angular position, and $\theta_{\rm res}=\lambda/D$ is expressed in arcseconds, with $\lambda=500$~nm and $D$ the circumscribed pupil diameter. Making an image brighter increases its gradient power even when spatial structure is unchanged. By dividing $\gabs$ by $\snrarc^2$, we reduce this increase in gradient power caused by brightness alone, although some brightness dependence remains due to relative pixel weights changing with source brightness.

Following the sample-selection cuts used by \citet{oriordan2023_euclid_sensitivity}, we retain systems with $\theta_{\rm E,eff}>0.5$~arcsec and $\snrarc>20$, leaving 965 systems. These are ranked by
\begin{equation}
\mathcal{R}=\mathcal{Z}(\log \snrarc)+\mathcal{Z}(\log \grel).
\label{eq:selection_score}
\end{equation}
Here, $\mathcal{Z}$ denotes the z-score, calculated by subtracting the mean and dividing by the population standard deviation across the 965 surviving systems. The ranking score therefore gives equal weight to the z-scores of log signal-to-noise ratio and log relative gradient power. The score was fixed before measuring injected-subhalo sensitivity, and in Section~\ref{sec:sensitivity}, we assess how well this ranking predicts subhalo sensitivity.

We assess the gain from target selection by comparing the mass reach of the 12, 25, 50, and 100 highest-ranked systems with that of the full ensemble of 965 survivors. Fisher sensitivity maps are calculated for all 965 systems (Section~\ref{subsec:sensitivity_maps}). The top 50 form the selected sample for the non-linear comparisons (Section~\ref{subsec:nonlinear_comparisons}), and the top 12 form the sample for the PSF tests in Section~\ref{sec:psf}. To test the Fisher forecasts beyond the highest-ranked systems, we select a pseudorandom validation sample of 100 systems from the full surviving ensemble, without using the selection score or measured subhalo sensitivity (Appendix~\ref{app:population_sampling}). Seven members also belong to the top 50, making up a total of 143 systems with non-linear fits. The ranking is fixed using only the expected subhalo-free images, intentionally set before evaluating subhalo sensitivity.

\subsection{Point spread function model}
\label{subsec:psf_model}

We generate the PSFs with the High Contrast Imaging in Python (HCIPy) package \citep{por2018_hcipy}, using the EAC1 pupil described in Section~\ref{subsec:observing}. The aberrated PSF combines both global Zernike modes over the pupil and hexagonal Zernike (hexike) modes within individual mirror segments. Hexikes form an orthonormal polynomial basis on a hexagonal pupil \citep{mahajan2006_hexikes}, and we contributed their segmented implementation into HCIPy for this project. The model uses global Noll indices 4--55 and the first ten segment modes, including piston, tip, and tilt. HCIPy then propagates the pupil wavefront to the focal plane at 500~nm. The intensity is integrated over the image pixels and each detector-sampled kernel is normalized to unit total flux. 

We derive the relative mode weights from publicly available optical path difference (OPD) maps produced by JWST wavefront sensing \citep{lajoie2023_jwst_wavefront}, retrieved from the Mikulski Archive for Space Telescopes (MAST) through the STPSF package\footnote{\url{https://stpsf.readthedocs.io/en/latest/jwst_measured_opds.html}}. We decompose each map into global Zernike modes over the full pupil, then fit hexike modes within each segment to the residual. Individual OPD maps determine the static weights, while differences between consecutive maps, excluding pairs that span a commanded mirror correction, determine the drift weights. The fitted JWST coefficients thus determine the weights of the individual modes and the balance between global and segment aberrations. Each mode coefficient is drawn from a zero-mean Gaussian distribution whose width is proportional to its derived weight. The combined wavefront is then rescaled to a desired RMS optical path difference over the illuminated EAC1 pupil.

The PSF-quality experiment varies the RMS amplitude of the reference wavefront, using the resulting PSF to both generate and model the images. For the PSF-knowledge experiment, the simulated images retain the reference PSF, corresponding to 35~nm RMS wavefront error, while an additional wavefront error is introduced into the fitted PSF. This error is generated using the drift weights and scaled to a prescribed RMS before being added to the reference wavefront; the amplitudes are chosen for this study rather than predicted for HWO. Section~\ref{sec:psf} presents the resulting changes in subhalo sensitivity.

\begin{figure*}
    \centering
    \includegraphics[width=\textwidth,alt={A nineteen-segment hexagonal pupil has spatially varying positive and negative optical path differences. The resulting point spread function has a bright central core surrounded by fainter, irregular diffraction structure. An inset enlarges the core.}]{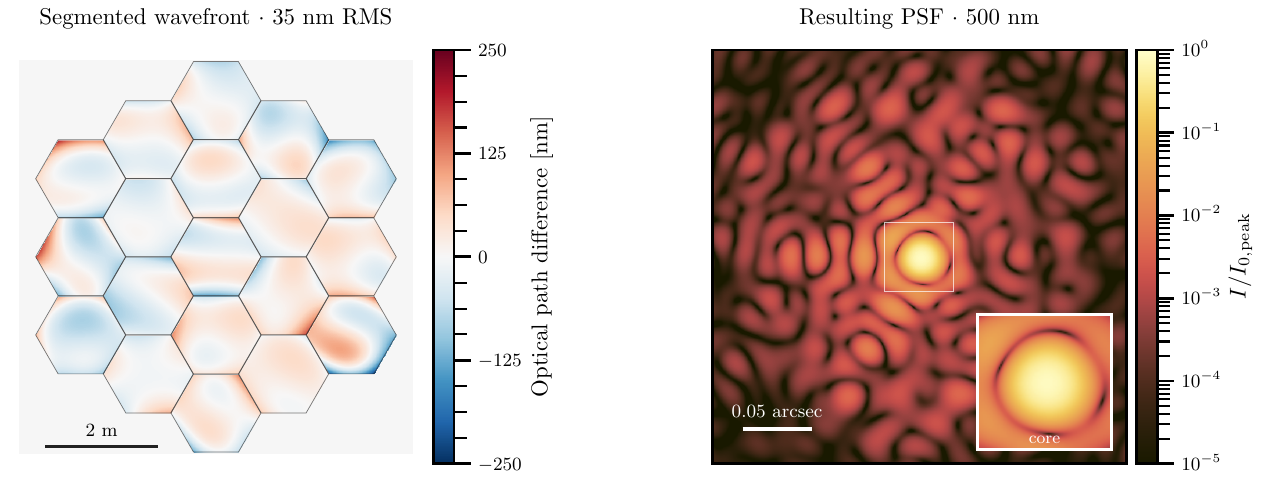}
    \caption{Reference wavefront and resulting PSF for the EAC1 pupil. Left: optical path difference across the segmented pupil, with the pupil mean removed and the wavefront scaled to 35~nm RMS. Right: the corresponding optical PSF at 500~nm, shown before detector-pixel integration. The intensity is normalized to the peak of the unaberrated PSF, $I_{0,\mathrm{peak}}$, and displayed on a logarithmic colour scale. The inset enlarges the PSF core.}
    \label{fig:psf_reference}
\end{figure*}

\section{Detection metrics}
\label{sec:metrics}

In this section, we define the metrics used to quantify subhalo detection sensitivity. To limit the computational cost, we use the Fisher approximation for sensitivity maps and reserve non-linear fits for targeted validation. We first describe the Fisher detection statistic (Section~\ref{subsec:fisher_statistic}), then explain how sensitivity maps give the mass thresholds used to compare lens systems (Section~\ref{subsec:sensitivity_maps}). Finally, we describe the non-linear fits used to test these forecasts (Section~\ref{subsec:nonlinear_comparisons}).

\subsection{Fisher detection statistic}
\label{subsec:fisher_statistic}

Small changes to the macro lens and source parameters can reproduce part of the image perturbation caused by a subhalo. Allowing these parameters to adjust is therefore necessary to avoid overstating the subhalo detection significance. Initially motivated by computational constraints, we develop a Fisher approximation to evaluate the residual subhalo signal across a wide set of subhalo masses and positions without a full non-linear search at each point.

The agreement between the observed and modelled pixels is quantified by the Gaussian log likelihood
\begin{equation}
\log\mathcal{L}(\boldsymbol{\eta}) = -\frac{1}{2}\sum_i
\frac{[d_i-m_i(\boldsymbol{\eta})]^2}{\sigma_i^2}+\mathrm{constant},
\label{eq:image_likelihood}
\end{equation}
where $d_i$ and $m_i$ are the observed and modelled electron counts in pixel $i$, $\boldsymbol{\eta}$ is the vector of lens, source, and background parameters, and $\sigma_i^2$ is the pixel variance.
The variances follow the noise model in Section~\ref{subsec:observing} with derived values from the parameters in Table~\ref{tab:observing_configuration}. These variances are fixed when comparing between smooth and subhalo models.

To determine whether the measurement favours a subhalo, we compare the best-fitting smooth model with a model that includes a subhalo. The improvement in fit is quantified by the likelihood-ratio statistic
\begin{equation}
q = 2\left(\log\widehat{\mathcal{L}}_{\mathrm{subhalo}}-\log\widehat{\mathcal{L}}_{\mathrm{smooth}}\right) = \chi^2_{\mathrm{smooth},\min}-\chi^2_{\mathrm{subhalo},\min},
\label{eq:likelihood_ratio}
\end{equation}
where the hats denote maximized likelihoods, and $\chi^2$ is the noise-weighted sum of squared image residuals. For the Fisher forecasts, the data $d_i$ are the expected electron counts in the simulated image containing a subhalo. These are referred to as Asimov data \citep{cowan2011_asymptotic}, with observational noise accounted for through the pixel variances $\sigma_i^2$ in the likelihood.

The Fisher calculation includes ten nuisance parameters: the lens centre, Einstein radius, and two ellipticity components; the source centre, flux normalization, and size; and a constant background offset. The source template's internal structure and rotation remain fixed, as does the PSF, and no priors are imposed on the nuisance shifts.

For each trial subhalo mass $M=M_{200}$ and lens-plane position $\boldsymbol{x}$, we calculate the PSF-convolved image containing the subhalo. Its difference from the corresponding smooth image is $\Delta m_i(M,\boldsymbol{x})$. The response of the smooth model to small parameter shifts is approximated by a first-order expansion about the input parameters $\boldsymbol{\eta}_0$:
\begin{equation}
m_{\mathrm{smooth},i}(\boldsymbol{\eta}_0+\delta\boldsymbol{\eta})
\simeq m_{\mathrm{smooth},i}(\boldsymbol{\eta}_0)+\sum_j J_{ij}\,\delta\eta_j,
\label{eq:linearized_smooth_model}
\end{equation}
where $\boldsymbol{\eta}_0$ contains the input smooth-model parameters, and $\delta\eta_j$ is a shift in nuisance parameter $j$. The image Jacobian $J$ has elements $J_{ij}=\partial m_{\mathrm{smooth},i}/\partial\eta_j$, evaluated at $\boldsymbol{\eta}_0$, describing how the modelled count in pixel $i$ changes with parameter $j$. The two terms of equation~\eqref{eq:likelihood_ratio} then simplify for different reasons. The input subhalo model reproduces the Asimov data exactly, so $\chi^2_{\mathrm{subhalo},\min}=0$. In the smooth term, equation~\eqref{eq:linearized_smooth_model} makes the residual linear in the nuisance parameter shifts. Minimizing that residual over the shifts gives the Fisher statistic $\qf$

\begin{equation}
\qf(M,\boldsymbol{x}) = \min_{\delta\boldsymbol{\eta}}\sum_i
\frac{\left[\Delta m_i(M,\boldsymbol{x})-\sum_j J_{ij}\,\delta\eta_j\right]^2}{\sigma_i^2}.
\label{eq:fisher_statistic}
\end{equation}
We evaluate the statistic over all pixels of the simulated image, allowing the nuisance parameters to adjust jointly in the minimization.

To evaluate equation~\eqref{eq:fisher_statistic}, we find the combination of nuisance parameter shifts that best reproduces the subhalo perturbation. This separates the perturbation into the part that the smooth model can absorb and the residual that remains. Because the residual is linear in the nuisance shifts, this minimization has a closed-form solution, avoiding a non-linear search over the parameters. The solution is determined by the nuisance Fisher matrix $F=J^{\mathsf T}C^{-1}J$, where $C=\mathrm{diag}(\sigma_i^2)$ is the image-noise covariance matrix. The Fisher matrix describes how strongly the image constrains different combinations of nuisance parameters. The best-fitting shifts are $\widehat{\delta\boldsymbol{\eta}}=F^{+}J^{\mathsf T}C^{-1}\Delta\boldsymbol{m}$, where $F^{+}$ is the pseudoinverse of $F$ and handles nuisance directions that leave the image unchanged. Substituting them into equation~\eqref{eq:fisher_statistic} gives
\begin{equation}
\qf = \Delta\boldsymbol{m}^{\mathsf T}C^{-1}\Delta\boldsymbol{m} - \widehat{\delta\boldsymbol{\eta}}^{\mathsf T}F\,\widehat{\delta\boldsymbol{\eta}}.
\label{eq:fisher_profiled_information}
\end{equation}
The first term measures the noise-weighted strength of the perturbation with the smooth model held fixed. The second subtracts the contribution reproduced by the best-fitting nuisance shifts, leaving the signal that remains unexplained within the linear approximation.

We adopt $\qf\geq10$, a log-likelihood gain of 5, following the detection threshold of the HWO strong-lensing science case \citep{he2026_hwo25_strong_lensing}. This statistic defines the sensitivity maps of Section~\ref{subsec:sensitivity_maps} and is tested against non-linear fits in Section~\ref{subsec:nonlinear_comparisons}.

For the PSF-modelling-error tests in Section~\ref{subsec:psf_modeling_errors}, the images are generated with the true PSF but modelled with an incorrect PSF. This mismatch can leave $\chi^2_{\mathrm{subhalo},\min}$ nonzero, so the assumption used to reduce the likelihood ratio to equation~\eqref{eq:fisher_statistic} no longer holds in general. We therefore retain both terms of equation~\eqref{eq:likelihood_ratio} and evaluate them using the linear approximation in equation~\eqref{eq:linearized_smooth_model}. Let $r_i=d_i-m_{\mathrm{smooth},i}$ be the residual between the true expected image and the smooth image calculated with the fitted PSF. We compute the nuisance responses $J_{ij}$ and the trial subhalo perturbation $\Delta m_i$ with the same fitted PSF. We fit a scale factor $a$ to adjust the strength of the trial subhalo perturbation when matching it to the residual image, giving
\begin{equation}
\begin{aligned}
q ={}& \min_{\delta\boldsymbol{\eta}}\sum_i
\frac{\left[r_i-\sum_j J_{ij}\,\delta\eta_j\right]^2}{\sigma_i^2} \\
&-\min_{a,\,\delta\boldsymbol{\eta}}\sum_i
\frac{\left[r_i-a\Delta m_i-\sum_j J_{ij}\,\delta\eta_j\right]^2}{\sigma_i^2}.
\end{aligned}
\label{eq:psf_mismatch_statistic}
\end{equation}
The first term fits the smooth model alone; the second also includes the scaled subhalo perturbation, with the nuisance parameters allowed to readjust. Either a positive or negative amplitude can improve the fit, but a negative amplitude reverses the predicted subhalo perturbation. We therefore require the best-fitting amplitude $\hat a>0$ as well as $q\geq10$. Using the true expected image with or without a subhalo as $d_i$ gives the injection or no-subhalo map, respectively. In images without a subhalo, positions satisfying these conditions count as false detections. With a correct PSF and an injected subhalo, $r_i=\Delta m_i$, the second term vanishes at $a=1$, and $q$ reduces to $\qf$ in equation~\eqref{eq:fisher_statistic}.

\subsection{Sensitivity maps and mass thresholds}
\label{subsec:sensitivity_maps}

Although subhalo mass reach is the standard strong lensing literature statistic, the mass of a subhalo does not determine its detectability alone. The lensing geometry and source morphology of a given system determine where a subhalo produces a measurable perturbation. We evaluate $\qf(M,\boldsymbol{x})$ on a grid of `trial' subhalo positions spaced by $\Delta x=0.05$~arcsec, within a $2\theta_{\rm E,eff}$ radius centred on the lens. The population has a range of area-equivalent Einstein radii $\theta_{\rm E,eff}$ of $0.5$ to $3.05$~arcsec, so we scale the aperture with $\theta_{\rm E,eff}$ so that the area fractions can be compared between lenses. This choice of aperture, including the factor of $2$, follows \citet{oriordan2023_euclid_sensitivity} who report sensitive-area fractions within $2\theta_{\rm E}$ for Euclid. Evaluating $\qf$ across these positions gives both the lowest mass a system can reach and how much of the aperture is sensitive at each mass.

\begin{figure*}
\centering
\includegraphics[width=\textwidth,clip,alt={Three panels show a lensed arc, its detectable regions at increasing subhalo masses, and the growth of sensitive aperture fraction with mass. The detectable region expands around the arc, with markers identifying first detection, ten per cent coverage, and fifty per cent coverage.}]{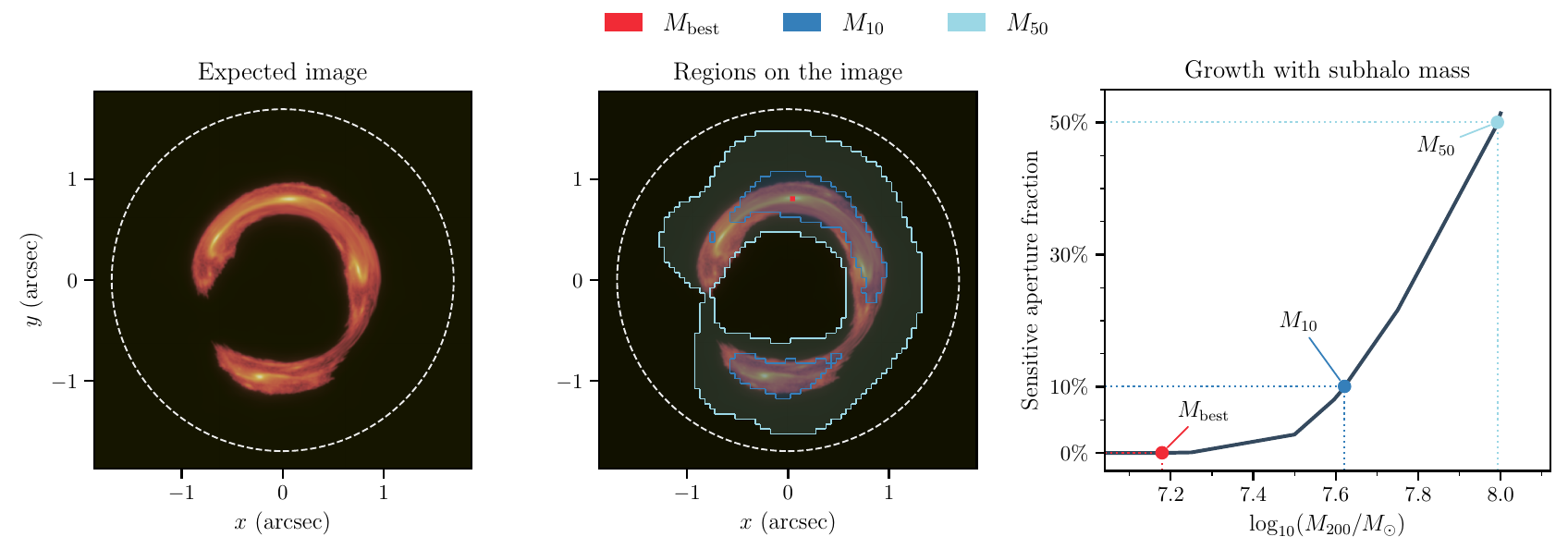}
\caption{Illustration of the subhalo mass thresholds for an example lens. Left: expected lensed image, with the dashed circle marking the aperture of radius $2\theta_{\rm E,eff}$. Middle: trial subhalo positions satisfying $\qf\geq10$ at $M_{\rm best}$ (red), $M_{10}$ (dark blue), and $M_{50}$ (light blue). Right: sensitive aperture fraction as a function of subhalo mass. The coloured markers indicate the mass at which the first trial position becomes detectable (red) and the masses at which sensitivity reaches 10\% and 50\% of the aperture (dark and light blue, respectively).}
\label{fig:sensitivity_regions}
\end{figure*}

At each mass, the sensitive area $A(M)$ is the total area of the trial cells satisfying $\qf\geq10$, and $f(M)$ is the fraction of cells that pass this threshold within the total aperture,
\begin{equation}
A(M) = N_{\rm det}(M)\,(\Delta x)^2,
\qquad
f(M) = \frac{N_{\rm det}(M)}{N_{\mathcal{A}}}.
\label{eq:sensitive_area}
\end{equation}
Here, $N_{\rm det}(M)$ is the number of trial positions inside the aperture with $\qf(M,\boldsymbol{x})\geq10$, and $N_{\mathcal{A}}$ is the total number of trial positions inside it. We repeat this calculation over $10^{6}\leq M_{200}/\msun\leq10^{9.5}$ in steps of 0.25~dex to estimate three mass thresholds for each system. $M_{\rm best}$ is the lowest mass at which $\qf$ reaches $10$ at any evaluated position. $M_{10}$ and $M_{50}$ are the masses at which $f(M)$ reaches 0.10 and 0.50, respectively, allowing us to distinguish between absolute low-mass reach of a system and one that is sensitive over a larger area. The angular area corresponding to 10\% aperture coverage is $0.1N_{\mathcal{A}}(\Delta x)^2\simeq0.1\pi(2\theta_{\rm E,eff})^2$, while the area for 50\% coverage is five times larger. Table~\ref{tab:aperture_areas} gives the median, mean, and range of the 10\% areas for the full ensemble and the selected samples. Figure~\ref{fig:sensitivity_regions} shows these three areas for a sample system. Additional 0.1-dex steps are added around the $M_{\rm best}$ crossing to achieve finer mass thresholding. Since the $\qf\geq10$ thresholds can fall between the sampled masses, we interpolate linearly in log mass between the two sampled masses on either side of the crossing. If the detection threshold is not reached by $10^{9.5}\msun$, we stop the search and record it as a lower bound, still retaining its place in the overall range for the computation of medians.

\subsection{Non-linear model comparisons}
\label{subsec:nonlinear_comparisons}

We test the Fisher forecasts with non-linear fits that allow finite changes to the lens and source parameters. Each image is fitted with PyAutoLens \citep{nightingale2021_pyautolens} using a smooth lensing system model and the same model with a subhalo. Both the SIE model family and the source template's internal structure and rotation are fixed, allowing a controlled comparison with the Fisher calculation. The centre, Einstein radius, and ellipticity of the lens are free, as are the centre, flux, and size of the source, within uniform priors centred on the input values (Table~\ref{tab:nonlinear_priors}). The widths are informed by the default prior-passing settings of the Source, Light and Mass (SLaM) pipelines in PyAutoLens, which are used by \citet{he2022_cosmo_subhalo_detection} and \citet{nightingale2023_subhalos}. The subhalo has the NFW profile and concentration defined in Section~\ref{subsec:lens_subhalo}, with a log-uniform mass prior between $10^{6}$ and $10^{11}\msun$. Its centre varies about the test position by $\pm0.15$~arcsec in each coordinate. 

\begin{table}
\centering
\caption{Priors of the non-linear fits. Each parameter is uniform within the listed half-width of its input value. The subhalo mass prior is uniform over the listed range.}
\label{tab:nonlinear_priors}
\small
\begin{tabular}{@{}ll@{}}
\toprule
Parameter & Prior half-width or range \\
\midrule
Lens centre & $\pm0.05$~arcsec \\
Einstein radius & $\pm25$ per cent \\
Lens ellipticity components & $\pm0.2$ \\
Source centre & $\pm0.05$~arcsec \\
Source flux & $\pm50$ per cent \\
Source size & $\pm50$ per cent \\
Subhalo centre & $\pm0.15$~arcsec of the test position \\
Subhalo mass & $6.0\leq\log_{10}(M_{200}/\msun)\leq11.0$ \\
\bottomrule
\end{tabular}
\end{table}

We use Nautilus \citep{lange2023_nautilus}, run through PyAutoFit \citep{nightingale2021_pyautofit}, to search the parameter space for each model. Since the highest-likelihood sample need not locate the maximum precisely, we refine each fit with bounded L-BFGS-B optimization \citep{byrd1995_lbfgsb} in SciPy \citep{virtanen2020_scipy} from multiple starting points. Twice the signed difference between the subhalo and smooth-model log-likelihood maxima gives $\qfit$ (equation~\eqref{eq:likelihood_ratio}), and we require $\qfit\geq10$ for detection. Before classifying a comparison, we check that both maxima are repeatable and agree with independent likelihood calculations. If either fit fails these checks, we repeat both searches once with more thorough sampling and apply the same refinement and checks. Comparisons that still fail are reported as unresolved, without a detection classification. The sampler settings and numerical checks are given in Appendix~\ref{app:nonlinear_numerics}.

We hold the supplied PSF fixed in each non-linear fit. The non-linear fits use a $51\times51$-pixel PSF kernel and exclude a border of half the kernel width, where the convolution is incomplete. With the $999\times999$-pixel kernel of the production Fisher maps, this border would cover most of each image. The fits also subtract the known background, whereas the Fisher maps allow a background offset to vary. We compare $\qfit$ with the production Fisher prediction at the same test mass and position. For the top 12, we also calculate the linearized statistic using the same kernel, fitted pixels, noise weights, and background treatment as the non-linear fits. This separates the effect of non-linear adjustment from the differences in image treatment.

We apply these fits to the 143 systems defined in Section~\ref{subsec:selection}. Asimov injections test the Fisher approximation using expected images, while noisy injections test recovery in individual noise realizations. We use no-subhalo controls to measure false detections. Each system has one noisy injection and one corresponding control. For the top 50, nine additional control realizations give ten controls per system and 500 in total.

For each system, the test mass is the upper sampled mass bracketing the correct-PSF $M_{\rm best}$ crossing in the production Fisher maps. Systems without a crossing are tested at the upper sampled limit of $10^{9.5}\msun$. At this mass, we recompute the Fisher map using the kernel of the non-linear fits and define the test position as the aperture position, clear of the excluded border, where the statistic is largest. The Asimov and noisy injections share this mass and position, and the control uses the same local search window without an injected subhalo. For the top 12, we also use the Asimov injection at the lower bracketing mass, with its position selected independently by the same rule at that mass. Together, the two masses give the 24 Asimov comparisons in Section~\ref{subsec:nonlinear_results}. The PSF-error tests in Section~\ref{subsec:nonlinear_psf_results} reuse the top-12 injection and control images while changing the supplied PSF. These non-linear comparisons test selected masses and positions, while the $M_{10}$ and $M_{50}$ thresholds are derived from the full Fisher sensitivity maps.

\section{Subhalo detection sensitivity}
\label{sec:sensitivity}

In this section, we present the subhalo detection sensitivity and mass reach of the lens population when the 35~nm RMS reference PSF is modelled correctly. We first describe the distribution of mass thresholds across the 965 surviving systems and the five source templates (Section~\ref{subsec:mass_sensitivity}), then examine how pre-detection selection by $\mathcal{R}$ changes the mass reach (Section~\ref{subsec:selection_effect}). Finally, we validate the Fisher forecasts against targeted non-linear fits (Section~\ref{subsec:nonlinear_results}).

\subsection{Subhalo mass sensitivity}
\label{subsec:mass_sensitivity}

\begin{figure*}
\centering
\includegraphics[width=\textwidth,clip,alt={Three cumulative distributions compare mass thresholds across source templates. Requiring greater aperture coverage shifts the distributions towards higher masses. Barred-spiral systems generally reach lower masses, while smooth-disc systems reach higher masses. Curves ending below one hundred per cent indicate unreached thresholds.}]{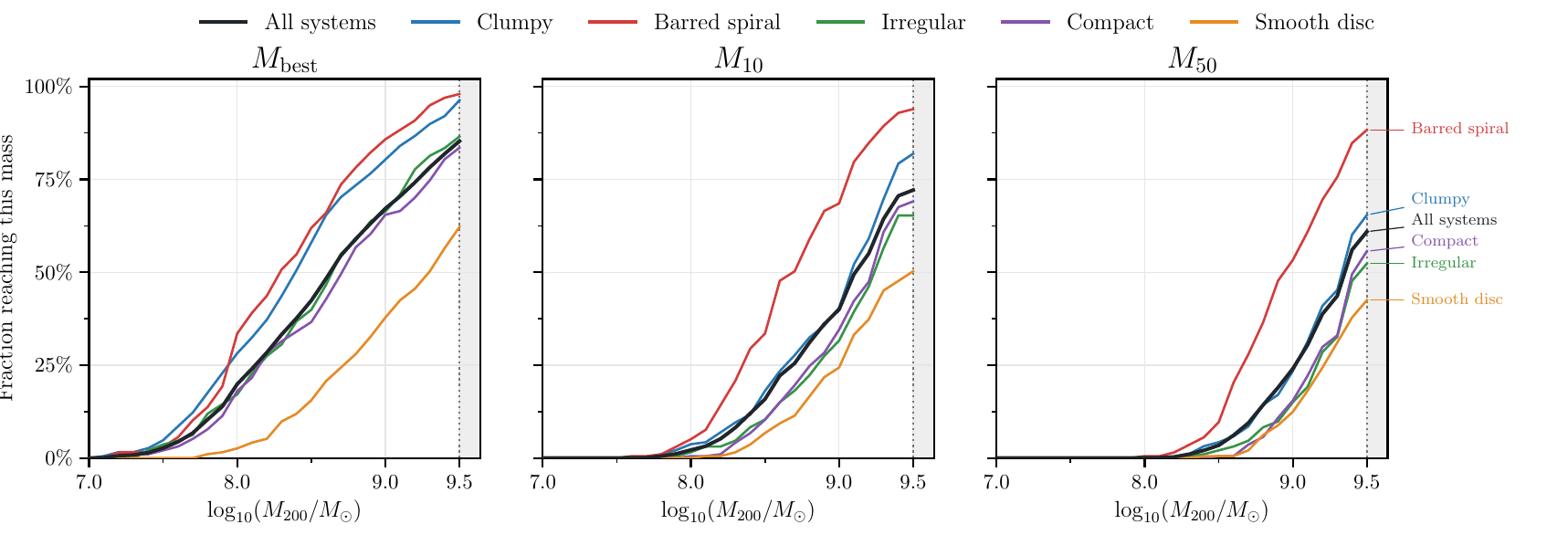}
\caption{Cumulative distributions of the three subhalo mass thresholds for the 965 surviving lens systems. From left to right, the panels show $M_{\rm best}$, $M_{10}$, and $M_{50}$. The black curves represent the full ensemble, while the coloured curves show systems grouped by source template. Each curve gives the fraction of systems whose threshold lies at or below a given mass. The dotted vertical line marks the upper sampled mass, $10^{9.5}\msun$, where the mass search stops. Systems whose thresholds are not reached within the sampled range remain in the denominators, so the curves end below 100\%.}
\label{fig:ensemble_mass_reach}
\end{figure*}

Figure~\ref{fig:ensemble_mass_reach} shows the cumulative distributions of $M_{\rm best}$, $M_{10}$, and $M_{50}$ for the 965 surviving systems and each source template. The lowest detectable threshold in any system is $M_{\rm best}=10^{7.02}\msun$, reached at a single position in a system using the clumpy source template. The median $M_{\rm best}$ of the entire ensemble is $10^{8.62}\msun$ (Table~\ref{tab:mass_reach}). Compared to the median $M_{\rm best}$, requiring the subhalo to be discoverable across a breadth of area raises the mass meaningfully. Requiring 10\% ($M_{10}$) of aperture coverage raises the median mass threshold to $10^{9.11}\msun$, and 50\% ($M_{50}$) of the aperture coverage raises the threshold to $10^{9.36}\msun$, corresponding to increases of approximately 0.49 and 0.74~dex, respectively. Within the sampled mass range, 142 systems cannot detect a subhalo at any evaluated position. For the area thresholds, 269 and 377 systems cannot detect a subhalo over 10\% and 50\% of the aperture, respectively. These systems remain in the distributions and median calculations.

For absolute mass reach ($M_{\rm best}$), the source templates fall into three general groups of strong, medium, and weak low-mass sensitivity. The barred-spiral and clumpy templates have the lowest medians, $10^{8.29}$ and $10^{8.40}\msun$, the irregular and compact templates track the ensemble median at $10^{8.65}$ and $10^{8.70}\msun$, while the smooth-disc template is highest at $10^{9.29}\msun$, a full dex above the barred spiral. $M_{10}$ and $M_{50}$ rearrange these groups, with the clumpy template moving down towards the ensemble median, while the barred spiral generally keeps the highest fraction. The smooth disc degrades even further: at $M_{50}$, more than half of its systems (111 of 193) do not reach 50\% aperture coverage, so its median lies above $10^{9.5}\msun$.

The median relative gradient power $\grel$ (equation~\ref{eq:selection_quantities}) of each template sample follows the same grouping as $M_{\rm best}$: 0.13 and 0.11 for the clumpy and barred-spiral samples, 0.10 and 0.08 for the irregular and compact samples, and 0.02 for the smooth disc. Each system's $\grel$ is measured from its expected lensed image used for ranking, so systems sharing a template often differ in $\grel$ due to their lens configuration and source placement. The ordering also does not directly translate to area medians: the clumpy sample has the highest median $\grel$, while the barred spiral has the lowest $M_{10}$ and $M_{50}$. As an aperture-wide summary, $\grel$ does not show where on the arc gradients lie. A possible explanation is that the clumpy template concentrates its brightest structure in the core and a few knots (Fig.~\ref{fig:source_lensing_gallery}), giving a low threshold at favourable positions without making a larger area sensitive, whereas the barred spiral has fine structure across most of its extent. \citet{despali2022_sensitivity} similarly found that a source with more complex, extended structure was sensitive over a larger region, not only near the brightest parts of the lensed arc. Each template represents only one galaxy of its morphology class, not a source morphology population, so these comparisons do not isolate the cause.

The template alone does not determine the mass reach of an individual system. Although the barred-spiral and smooth-disc medians differ by a full dex, 17 of the 193 smooth-disc systems reach a lower $M_{\rm best}$ than the barred-spiral median. Section~\ref{subsec:selection_effect} examines how ranking systems by $\mathcal{R}$ identifies the configurations with the lowest thresholds.

\begin{table}
\centering
\caption{Median subhalo mass thresholds for samples selected by the pre-detection ranking $\mathcal{R}$, the full surviving ensemble, and groups defined by source template. $N$ is the number of systems in each sample. Thresholds not reached within the sampled mass range are treated as lying above $10^{9.5}\msun$ when calculating the medians. The smooth-disc $M_{50}$ entry of ${>}9.5$ indicates that more than half of the group has an unreached threshold. The unreached fraction for each group is one minus the corresponding curve endpoint in Fig.~\ref{fig:ensemble_mass_reach}. No threshold is unreached in the ranked samples.}
\label{tab:mass_reach}
\small
\setlength{\tabcolsep}{6pt}
\renewcommand{\arraystretch}{1.08}
\begin{tabular*}{\columnwidth}{@{\extracolsep{\fill}}lrrrr@{}}
\toprule
Sample & $N$ & \multicolumn{3}{c}{Median $\log_{10}(M_{200}/\msun)$} \\
\cmidrule(l){3-5}
 & & $M_{\rm best}$ & $M_{10}$ & $M_{50}$ \\
\midrule
\multicolumn{5}{@{}l}{\textit{Ranked by $\mathcal{R}$}} \\
Top 12 & 12 & 7.19 & 7.82 & 8.30 \\
Top 25 & 25 & 7.42 & 8.00 & 8.45 \\
Top 50 & 50 & 7.51 & 8.12 & 8.54 \\
Top 100 & 100 & 7.67 & 8.23 & 8.63 \\
\addlinespace[0.6ex]
All systems & 965 & 8.62 & 9.11 & 9.36 \\
\addlinespace[0.8ex]
\multicolumn{5}{@{}l}{\textit{By source template}} \\
Barred spiral & 197 & 8.29 & 8.68 & 8.94 \\
Clumpy & 188 & 8.40 & 9.06 & 9.33 \\
Irregular & 193 & 8.65 & 9.26 & 9.44 \\
Compact & 194 & 8.70 & 9.22 & 9.41 \\
Smooth disc & 193 & 9.29 & 9.48 & ${>}9.5$ \\
\bottomrule
\end{tabular*}
\end{table}

\subsection{The effect of target selection}
\label{subsec:selection_effect}

\begin{figure*}
\centering
\includegraphics[width=\textwidth,clip,alt={Three scatter plots show mass thresholds generally rising as numerical rank increases, placing the most sensitive systems towards the left. Diamonds mark the highest-ranked system and stars mark the lowest thresholds. Upward triangles indicate thresholds beyond the sampled mass range.}]{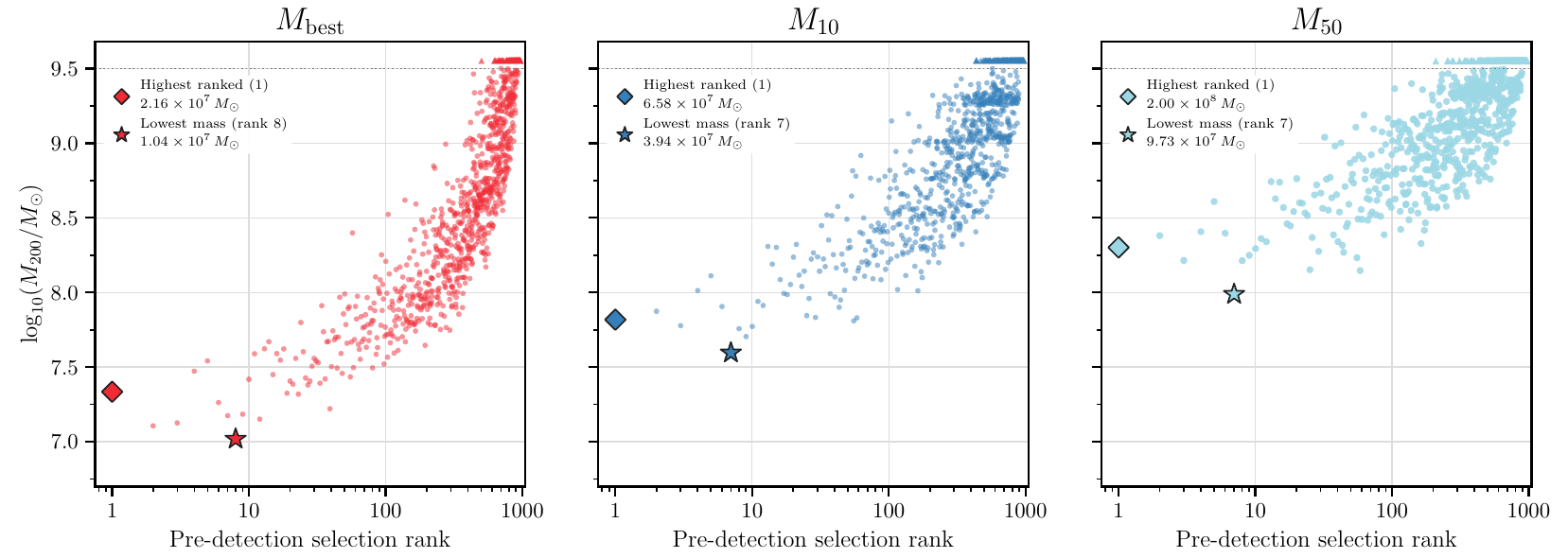}
\caption{Subhalo mass thresholds as a function of pre-detection selection rank by $\mathcal{R}$ for the 965-system ensemble. From left to right, the panels show $M_{\rm best}$, $M_{10}$, and $M_{50}$. Each point represents one system, with rank 1 corresponding to the highest $\mathcal{R}$ score, shown on a logarithmic axis. Diamonds mark the highest-ranked system, while stars identify the system with the lowest mass threshold in each panel. The dotted horizontal line marks the upper sampled mass, $10^{9.5}\msun$. Upward triangles indicate systems that do not reach the corresponding sensitivity threshold within the sampled mass range.}
\label{fig:selection_rank_mass_reach}
\end{figure*}

The ranking statistic $\mathcal{R}$ was designed in the spirit of a follow-up programme that selects targets before observation and subsequent subhalo detection (Section~\ref{subsec:selection}). Figure~\ref{fig:selection_rank_mass_reach} shows that the mass thresholds generally increase with pre-detection rank, indicating that $\mathcal{R}$, combining $\snrarc$ and $\grel$, identifies configurations with lower thresholds. Across all 965 systems, the Spearman correlations between numerical rank and $M_{\rm best}$, $M_{10}$, and $M_{50}$ are 0.93, 0.86, and 0.82, respectively (systems with thresholds unreached at $10^{9.5}\msun$ are placed after all measured thresholds, tied with one another). Within each source template, the $M_{\rm best}$ correlation is 0.90--0.92, so the ranking also orders configurations that share a template. The lowest value of each threshold occurs within the top ten, though not at pre-detection rank 1, and all of the 200 highest-ranked systems reach all three thresholds.

For the top 50 systems, the median thresholds are $M_{\rm best}=10^{7.51}\msun$, $M_{10}=10^{8.12}\msun$, and $M_{50}=10^{8.54}\msun$ (Table~\ref{tab:mass_reach}), which are approximately 1.1, 1.0, and 0.8~dex below the ensemble medians, respectively. For $M_{10}$, this top-50 pre-detection selection results in a factor of ten better mass reach. The top 12, which form the sample for the PSF tests in Section~\ref{sec:psf}, reach further, with medians of $M_{\rm best}=10^{7.19}\msun$, $M_{10}=10^{7.82}\msun$, and $M_{50}=10^{8.30}\msun$. The gain declines slowly with sample size, as seen in Table~\ref{tab:mass_reach}. The top 50 contains no smooth-disc systems, but 19 of its members use the irregular and compact templates, whose ensemble medians track the full population (Section~\ref{subsec:mass_sensitivity}). The ranking therefore selects favourable configurations within these templates as well as the templates with the lowest thresholds. We do not propose $\mathcal{R}$ as the statistic a future HWO programme would use; it stands in for selecting the most favourable systems from a larger sample. Because it is computed from the noise-free expected images, the selection gain realized by a future programme depends on how well these systems can be identified from noisy survey images. The top 50 forms the selected sample for the non-linear tests in Section~\ref{subsec:nonlinear_results}.

\subsection{Comparison with non-linear fits}
\label{subsec:nonlinear_results}

\begin{figure*}
\centering
\includegraphics[width=0.8\textwidth,clip,alt={Blue points for the fifty selected systems lie close to equality between Fisher and non-linear detection statistics and above both detection thresholds. Grey validation points show greater downward scatter. Dashed threshold lines bound the shaded region where both methods detect the injected subhalo.}]{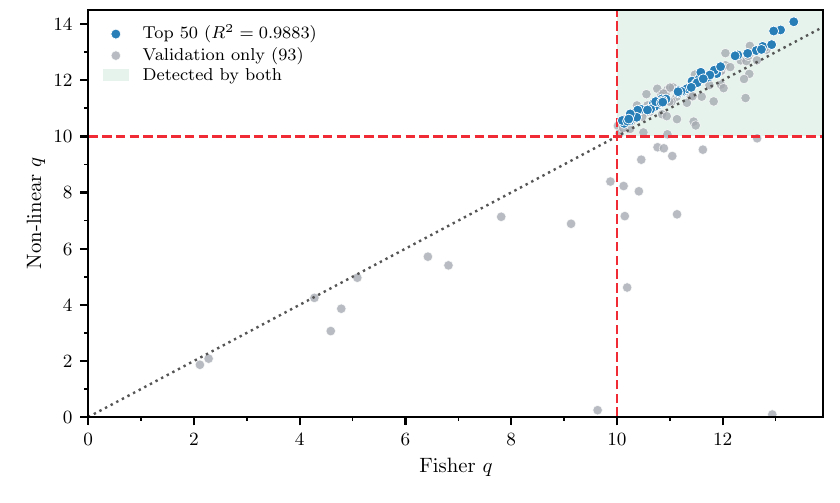}
\caption{Comparison of the Fisher statistic $\qf$ with the non-linear likelihood-ratio statistic $\qfit$ for Asimov injections in 143 lens systems. Blue points show the 50 highest-ranked systems, while grey points show the 93 validation systems outside this selection. Each system is evaluated at the upper test mass described in Section~\ref{subsec:nonlinear_comparisons}; systems without a Fisher threshold crossing are tested at $10^{9.5}\msun$. The dotted line marks equality, the dashed lines mark the detection threshold of $q=10$, and the shaded region indicates detections by both methods at $q\geq10$. The reported $R^2$ is the coefficient of determination of a linear fit to the top-50 points.}
\label{fig:asimov_validation}
\end{figure*}

Figure~\ref{fig:asimov_validation} shows the agreement between the Fisher and non-linear statistics for the top 50 systems and a randomly drawn validation sample of 100 systems. Seven systems drawn from the 965-system population overlap with the top 50, giving 143 distinct systems in total. All 50 Asimov injections in the selected sample satisfy $\qfit\geq10$, with $R^2=0.988$ for a linear fit between the two statistics. The median difference $\qfit-\qf$ is $+0.41$ relative to the production maps. Most of this offset is accounted for by the likelihood conventions of the non-linear fits (Section~\ref{subsec:nonlinear_comparisons}). For the top 12, a linear calculation using the same kernel, border, and background treatment as the fits lies a median of 0.39 above the production prediction and differs from $\qfit$ by less than 0.043 across the 24 upper- and lower-mass Asimov tests. The linear approximation therefore holds at these masses and positions, and the non-linear adjustment of the lens and source changes $q$ very little.

The validation sample shows where this tight agreement degrades. Of the 81 validation-only systems that the Fisher calculation predicts to be detectable, 69 are detected in the non-linear Asimov fits. All 12 misses occur at masses between $10^9$ and $10^{9.5}\msun$, above every top-50 test mass. These discrepancies are consistent with larger adjustments of the lens and source absorbing more of the perturbation than the linear approximation in the Fisher calculation allows. \citet{despali2022_sensitivity} similarly found that holding the main-lens parameters fixed gives overly optimistic sensitivity predictions for NFW perturbers (see their Appendix A); here, the Fisher calculation allows readjustment only through the first-order response. The threshold disagreements in Fig.~\ref{fig:asimov_validation} run in one direction: no system is detected by the non-linear Asimov fit but not by Fisher. The selection ranking that improves mass reach (Section~\ref{subsec:selection_effect}) also identifies systems for which the two statistics agree.

When changing from Asimov to single noise realizations, the injections are detected in 38 of the 50 selected systems, and in 58 of the 81 validation-only systems that are predicted to be detectable (Table~\ref{tab:nonlinear_outcomes}). For systems with a Fisher threshold crossing, the test mass lies just above the $M_{\rm best}$ crossing, most injections sit close to the detection threshold, and one noise realization can move an otherwise detectable injection below it. To test for false positives, we also fit noisy images of the same systems without a subhalo, using the same search around the same test position, and count any fit that reaches $\qfit\geq10$ as a false detection. With one noise realization per system, only 1 of the 49 accepted top-50 fits and none of the 93 accepted validation-only fits give a false detection. When we generate ten noise realizations for each of the top 50 systems, only 3 of 497 accepted fits give a false detection, each in a different system. The remaining 3 of the 500 fits did not pass the numerical checks and remain unclassified. These correct-PSF false-positive tests provide the baseline for the PSF-error tests in Section~\ref{sec:psf}.

\begin{table}
\centering
\caption{Non-linear detection and control outcomes with the correct PSF. $N$ is the number of systems. Validation only lists the 93 systems of the validation sample (Section~\ref{subsec:selection}) outside the top 50; the seven shared systems of the random 100-system draw appear in the first row. Injected detections count accepted non-linear fits with $\qfit\geq10$ among systems whose production Fisher statistic reaches $\qf\geq10$ at the tested mass and position. The 12 validation systems that do not reach this threshold within the sampled mass range were tested at $10^{9.5}\msun$ and are excluded from these denominators. Control exceedances (false detections) count accepted no-subhalo fits with $\qfit\geq10$ over all accepted no-subhalo fits, with unresolved fits in parentheses. The 10-draw column combines each top-50 system's single control with nine further noise realizations.}
\label{tab:nonlinear_outcomes}
\small
\setlength{\tabcolsep}{4pt}
\begin{tabular}{@{}lrcccc@{}}
\toprule
 & & \multicolumn{2}{c}{Injected detections} & \multicolumn{2}{c}{Control exceedances} \\
\cmidrule(lr){3-4}\cmidrule(l){5-6}
Sample & $N$ & Asimov & Noisy & 1 draw & 10 draws \\
\midrule
Top 50 by $\mathcal{R}$ & 50 & 50/50 & 38/50 & 1/49 (1) & 3/497 (3) \\
Validation only & 93 & 69/81 & 58/81 & 0/93 & -- \\
\bottomrule
\end{tabular}
\end{table}

\section{PSF degradation and mismodelling}
\label{sec:psf}

Section~\ref{sec:sensitivity} used the 35~nm RMS reference PSF and assumed that it is perfectly modelled. In this section, we relax that assumption and examine how PSF quality and modelling errors affect the 12 highest-ranked systems selected in Section~\ref{subsec:selection_effect}. We first vary the physical PSF while modelling it correctly (Section~\ref{subsec:psf_degradation}), then keep the true PSF fixed and introduce modelling errors in the fitted PSF (Section~\ref{subsec:psf_modeling_errors}). Finally, we test how these errors change subhalo recovery and false detections in the non-linear fits (Section~\ref{subsec:nonlinear_psf_results}).

\subsection{The effect of PSF degradation}
\label{subsec:psf_degradation}

\begin{figure*}
\centering
\includegraphics[width=\textwidth,clip,alt={Three line plots show mass thresholds increasing gradually with physical wavefront error. Thick coloured median curves rise less than the spread between individual grey system curves. Vertical dotted lines mark the reference wavefront error of thirty-five nanometres.}]{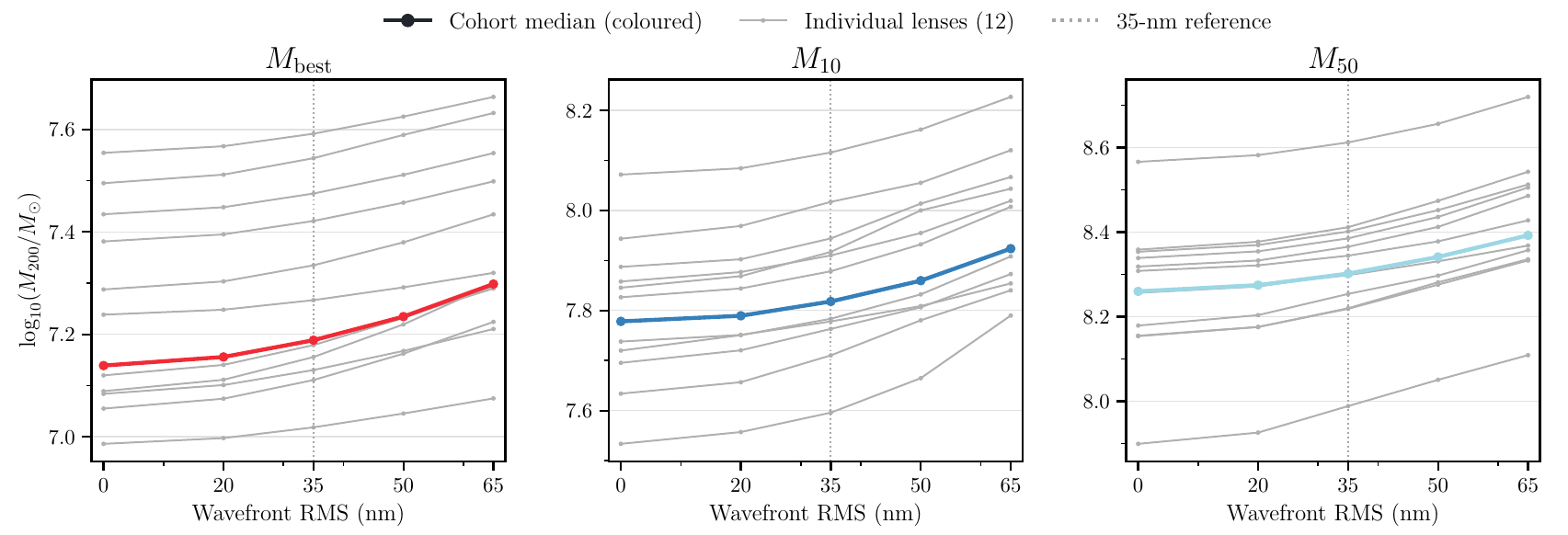}
\caption{Effect of PSF degradation on the Fisher mass thresholds for the 12 systems ranked highest by the pre-detection score $\mathcal{R}$. From left to right, the panels show $M_{\rm best}$, $M_{10}$, and $M_{50}$ as functions of RMS wavefront error. The same wavefront pattern is scaled to 0, 20, 35, 50, and 65~nm RMS. Thin grey curves show individual systems, while the thick coloured curves show the sample medians. The dotted vertical lines mark the 35-nm reference wavefront error adopted for the results in Section~\ref{sec:sensitivity}.}
\label{fig:psf_optical_quality}
\end{figure*}

Figure~\ref{fig:psf_optical_quality} shows how the three mass thresholds change when the reference wavefront (Fig.~\ref{fig:psf_reference}) is scaled from 0 to 65~nm RMS, with the resulting PSF used both to generate and to model each image. The sampled amplitudes span an unaberrated PSF at 0~nm, an optimistic HWO case at 20~nm, the 35-nm reference, a degraded case at 50~nm, and 65~nm, comparable to the in-flight wavefront error of JWST \citep{lajoie2023_jwst_wavefront}. The 35-nm reference meets the Maréchal criterion for diffraction-limited imaging at 500~nm (Strehl ratio $\geq0.8$).

The 35-nm RMS point corresponds to the top-12 row of Table~\ref{tab:mass_reach}. Relative to this reference, an unaberrated PSF lowers the median $M_{\rm best}$, $M_{10}$, and $M_{50}$ by only 0.05, 0.04, and 0.04~dex, respectively, while increasing the wavefront error to 65~nm raises them by 0.11, 0.11, and 0.09~dex. The Strehl ratio falls from 1 at 0~nm to 0.84 at 35~nm and 0.57 at 65~nm as aberrations move light out of the PSF core and reduce the contrast of fine image structure in the lensing system. Across this wavefront-error range, the changes in mass reach are substantially smaller than the variation among the 12 systems and the approximately 1-dex gain from target selection in Section~\ref{subsec:selection_effect}.

\subsection{The effect of PSF modelling errors}
\label{subsec:psf_modeling_errors}

To test how PSF knowledge affects subhalo sensitivity, we use the 35-nm reference PSF in the simulated images and add a drift-shaped wavefront error to the fitted PSF, following Section~\ref{subsec:psf_model}. For each of the 12 systems, we draw eight error patterns, referred to as directions, and scale each to 1, 2, 5, 10, and 20~nm RMS. We evaluate sensitivity at the upper sampled mass bracketing each system's correct-PSF $M_{\rm best}$, $M_{10}$, and $M_{50}$ thresholds. We compute the sensitivity maps with equation~\eqref{eq:psf_mismatch_statistic}, which measures how strongly the remaining image residuals resemble a subhalo after allowing first-order adjustments to the nuisance parameters.

\begin{figure*}
\centering
\includegraphics[width=\textwidth,clip,alt={Three maps overlay detectable regions on the same lensed image, using the correct point spread function and two modelling-error amplitudes. Blue and red hatching distinguish correct and mismodelled detections. Increasing error removes some originally detectable positions while adding regions elsewhere.}]{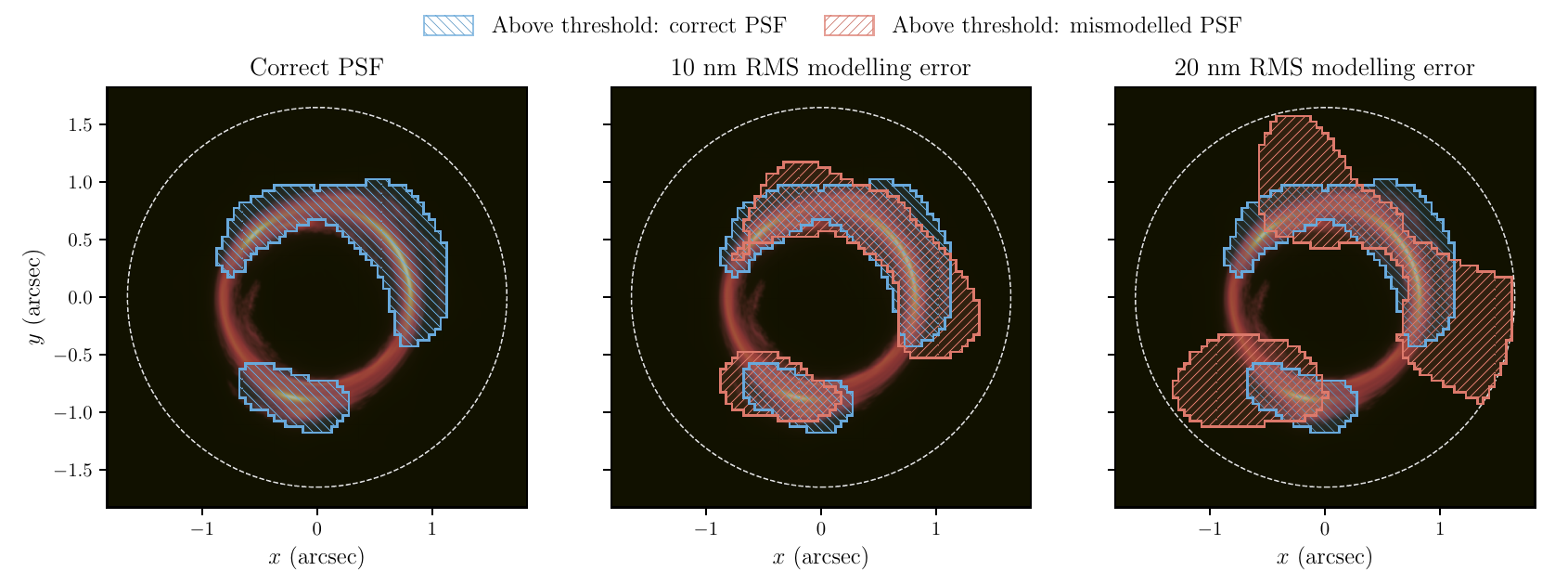}
\caption{Effect of PSF modelling errors on the highest-ranked system for an injected subhalo mass of $10^8\msun$. The background shows the Asimov lensed image in orange. Blue and red hatching mark positions passing the detection criterion in the injected maps using the correct and mismodelled PSFs, respectively. Overlap between the two denotes retained positions. This set of mismodelled detections in red represents the largest loss of original positions among the eight PSF directions. The dashed circle marks the aperture of radius $2\theta_{\rm E,eff}$.}
\label{fig:psf_mismodeling_regions}
\end{figure*}

PSF modelling errors can suppress or strengthen the fitted subhalo signal, changing which positions on the lens are sensitive. This is shown in Fig.~\ref{fig:psf_mismodeling_regions}. For the highest-ranked system with an injected subhalo mass of $10^8\msun$, the displayed 10-nm error leaves only 79\% of the originally sensitive positions above the threshold, while increasing the total sensitive area by 11\%. The corresponding no-subhalo map remains below $q=10$ throughout the aperture. The additional crossings occur only in the injected maps and indicate an increased apparent preference for a subhalo under the incorrect PSF.

To measure how much of the original sensitivity survives, we define $K$ as the fraction of correct-PSF sensitive positions that remain above threshold with the incorrect PSF. We also define $F$ as the area satisfying $\hat a>0$ and $q\geq10$ in the no-subhalo map, divided by the correct-PSF sensitive area at the same mass. Both quantities are computed from the expected images. As illustrative performance criteria, we require $Q_{10}(K)\geq0.9$ and $Q_{90}(F)\leq0.1$, where $Q_p$ is the $p$th percentile over the eight directions. At these percentiles, we require at least 90\% of the originally sensitive area to remain detectable, while limiting the area producing false detections to 10\% of that same original area. We define each system's tolerance as the largest sampled error amplitude satisfying both criteria, and apply these criteria at the $M_{10}$ and $M_{50}$ test masses.

Under these criteria, all 12 systems pass at 2~nm and most at 5~nm (Table~\ref{tab:psf_knowledge}). At 10~nm, only four systems pass at the $M_{10}$ test masses and two at the $M_{50}$ test masses, and none pass at 20~nm. Most systems therefore have a tolerance of 5~nm: seven of the 12 in the $M_{10}$ tests and eight in the $M_{50}$ tests pass at 5~nm and fail at 10~nm. Every system satisfies the $F$ criterion at every sampled amplitude, so loss of the original sensitive positions sets the tolerances.

\begin{table}
\centering
\caption{Effect of PSF modelling errors on the 12 highest-ranked systems. Error is RMS wavefront mismatch in nm. Fisher columns give the median $Q_{10}(K)$ across systems (Retained) and the number satisfying the criteria in Section~\ref{subsec:psf_modeling_errors} (Pass), using eight directions per system. Non-linear columns give the noisy-injection detections (Det.) and no-subhalo control false positives (Exc.) at $\qfit\geq10$, using three directions per system. Denominators include only accepted comparisons and parentheses list unresolved controls.}
\label{tab:psf_knowledge}
\small
\setlength{\tabcolsep}{3pt}
\begin{tabular}{@{}l*{6}{c}@{}}
\toprule
 & \multicolumn{4}{c}{Fisher maps} & \multicolumn{2}{c}{Non-linear fits} \\
\cmidrule(lr){2-5}\cmidrule(l){6-7}
 & \multicolumn{2}{c}{$M_{10}$} & \multicolumn{2}{c}{$M_{50}$} & Injections & Controls \\
\cmidrule(lr){2-3}\cmidrule(lr){4-5}\cmidrule(lr){6-6}\cmidrule(l){7-7}
Error (nm) & Retained & Pass & Retained & Pass & Det. & Exc. \\
\midrule
Correct PSF & -- & -- & -- & -- & 11/12 & 0/12 \\
2 & 0.97 & 12/12 & 0.97 & 12/12 & 33/36 & 0/35 (1) \\
5 & 0.92 & 11/12 & 0.92 & 10/12 & 32/36 & 0/36 \\
10 & 0.82 & 4/12 & 0.84 & 2/12 & 32/36 & 0/36 \\
20 & 0.67 & 0/12 & 0.66 & 0/12 & 28/36 & 1/36 \\
\bottomrule
\end{tabular}
\end{table}

\subsection{Non-linear fits with PSF modelling errors}
\label{subsec:nonlinear_psf_results}

We test whether the response to PSF modelling errors in the Fisher maps persists in non-linear fits to noisy images of the 12 highest-ranked systems. Three of the eight error directions are tested at 2, 5, 10, and 20~nm RMS, giving 36 conditions per amplitude. Each injection and control image is held fixed relative to its correct-PSF counterpart in Section~\ref{subsec:nonlinear_results}, so changes within a pair must arise from the PSF used in fitting. Only one condition detected with the correct PSF is lost for errors up to 10~nm, first falling below threshold at 5~nm. Increasing the error to 20~nm leaves 28 of the 33 initially detected conditions, with the five losses confined to two systems (Table~\ref{tab:psf_knowledge}). Every non-linear loss also falls below threshold in the Fisher map at the same test mass and position. At 20~nm, the Fisher calculation predicts 11 lost detections, compared with five in the non-linear fits. For the 11 systems detected with the correct PSF, the median Fisher statistic is 10.8, just above the detection threshold of 10, whereas the median statistic from the noisy non-linear fits is 14.0. The Fisher predictions therefore require a smaller decrease to fall below threshold, which may help explain the additional predicted losses.

PSF modelling errors also increase $\qfit$ in 22--24 of the 36 injected conditions at each amplitude. At 20~nm, changes relative to the correct-PSF fits span $-8.8$ to $+24.7$. These increases represent a greater apparent preference for the subhalo model but do not produce any newly detected injections in these tests.

None of the 12 correct-PSF controls produces a false detection, and no accepted control fit crosses the threshold for PSF errors up to 10~nm. A false detection appears in one of the 36 controls when the error reaches 20~nm. For this control and error direction, increasing the PSF error on the same noisy image raises $\qfit$ from 4.0 to 11.2. The non-linear tests therefore support the picture from the sensitivity maps, with PSF modelling errors causing missed detections while false detections remain rare within the tested search windows.

\section{Discussion}
\label{sec:discussion}

The HWO strong-lensing science case of \citet{he2026_hwo25_strong_lensing} identifies sensitivity to dark matter haloes near $10^7\msun$ as a route to testing alternatives to CDM. The study considers detection programmes of 500 to 2000 lensing systems. Wide-field surveys will discover far more lenses than such a programme can observe, with approximately 170,000 forecast for Euclid \citep{collett2015_euclid_predict, walmsley2026_euclid_dr0} and 160,000 for Roman's HLWAS \citep{wedig2025_roman_lenses}. The reach of a programme therefore depends on how well the most sensitive systems can be pre-selected for deep follow-up. Under common observing conditions, the 50 highest-ranked systems in our ensemble have a median $M_{\rm best}$ of $10^{7.51}\msun$, about 1.1~dex below the ensemble median, and the median for the 12 highest-ranked systems is $10^{7.19}\msun$. For scale, our top-50 and top-12 samples represent 5.2\% and 1.2\% of the simulated ensemble. Selecting the same fractions of the forecast Euclid or Roman populations would give samples of approximately 8,000--9,000 and 2,000 lenses, respectively. How many of the forecasted survey lenses reach these thresholds depends on various lens property factors that our population does not reproduce, such as the redshift, brightness, Einstein radius, and source morphology distributions. The variation in sensitivity across lens configurations therefore motivates identifying those most suited to low-mass subhalo searches before allocating HWO programme time.

For a future HWO follow-up programme, the aim would therefore be to identify promising targets from the larger survey population before following up and calculating their subhalo sensitivity. Pre-selecting the top 50 by $\mathcal{R}$ lowered the median $M_{10}$ from $10^{9.11}$ to $10^{8.12}\msun$ in our analysis, so the gain extends to the area over which subhaloes are detectable as well as the aforementioned $M_{\rm best}$ thresholds. \citet{oriordan2023_euclid_sensitivity} demonstrated the value of selecting lenses by the depth or area of their sensitivity maps, while our score uses arc signal-to-noise and relative gradient power measured directly from the expected images. Applying it to survey candidates would require estimating these quantities from noisy images with lower resolution than the HWO follow-up data will provide, and the realized gain will depend on how well the informative structure can be identified in these images. 

Full non-linear fits become computationally expensive when repeated across many lens systems, subhalo masses, positions, and observational configurations. Our comparisons show that the faster Fisher approximation closely reproduces the non-linear detection statistics at the masses and positions tested in the selected sample. All 50 selected systems are detected in the Asimov fits at their upper test masses and chosen positions. For the top 12, a linear calculation using the same PSF kernel, pixels, and background treatment as the non-linear fits gives Fisher statistics that differ from the non-linear values by less than 0.043. The threshold disagreements in the validation sample occur at $10^9$--$10^{9.5}\msun$, above every test mass in the selected sample, consistent with the larger lens and source adjustments absorbing more of the perturbation than the linear approximation in the Fisher method predicts. Fisher maps are therefore well suited for larger sweeps of observing configurations at low subhalo masses, with non-linear fits confirming these forecasts at selected positions and masses. Most injections are tested just above the Fisher detection threshold, so noise can push some just below it.

The PSF experiments in this study respond to \citet{he2026_hwo25_strong_lensing}, who identified precise PSF knowledge as essential for this science case and left the required PSF stability for future work. We test this by analysing how wavefront error affects the information in the image and how accurately the PSF must be modelled to recover it. For the top 12 systems, degrading the physical wavefront error from 35 to 65~nm RMS raises the median mass thresholds by about 0.1~dex along the tested wavefront pattern when the PSF is modelled correctly. \citet{wedig2025_roman_lenses} also found a modest effect in deep Roman imaging, where with detector sensitivity held fixed, the mean minimum subhalo mass detectable at $3\sigma$ varies by about 5\% across the focal plane and tracks the wavefront error. Their idealized $\chi^2$ statistic attributes every image difference to the subhalo, without refitting the lens or source. An incorrect PSF model can also introduce residuals that resemble a subhalo perturbation or cancel part of the fitted subhalo signal. This cancellation removes one of the 33 correct-PSF detections at 5 and 10~nm RMS, and five at 20~nm. Positions can also cross the detection threshold without any improvement in the observed image. Changing the fitted PSF can increase the apparent preference for a subhalo, although false detections remain rare in these tests. Only one of the 143 accepted PSF-error control comparisons crosses the detection threshold, at 20~nm RMS, where the statistic rises from 4.0 with the correct PSF to 11.2 on the same noisy image.

In the Fisher maps, loss of originally detectable positions sets the PSF-modelling tolerance. Under the illustrative criteria in Section~\ref{subsec:psf_modeling_errors}, the median tolerance is 5~nm RMS, with a range of 2--10~nm across the 12 systems for both the $M_{10}$ and $M_{50}$ tests. The tolerances describe the mismatch between the true and assumed wavefronts in the model, with no time-scale attached. Our PSF-error patterns use relative mode weights derived from differences between consecutive JWST OPD measurements. \citet{lajoie2023_jwst_wavefront} report a median measured change of 9~nm RMS over 48~hours during Cycle~1, including sensing noise estimated at 5--7~nm RMS. Our median tolerance of 5~nm RMS is comparable to that sensing noise and smaller than the median measured two-day change.

We generate and fit each image using the same source template and family of SIE lens mass models where the lens and source parameters can adjust, but the source's internal structure remains fixed. In observations, all structure must be inferred from the data, which can give the source model more freedom to absorb part of a subhalo perturbation. \citet{he2022_cosmo_subhalo_detection} showed that this additional freedom can reduce both subhalo sensitivity and false positives. The lens mass model can also leave residuals that resemble a subhalo if it cannot reproduce the underlying mass distribution, leading to false detections or biased masses, as shown by \citet{he2022_cosmo_subhalo_detection} and \citet{nightingale2023_subhalos}. We also omit lens-galaxy light, as in the simulated lenses of \citet{he2022_cosmo_subhalo_detection}, and \citet{despali2022_sensitivity} found that including correctly modelled lens light did not alter their sensitivity results. These effects can interact with PSF errors because each changes the residuals that the subhalo model fits. These forecasts are therefore optimistic with respect to source reconstruction and foreground light.

Our adopted concentrations are lower than the \cite{moline2017_subhalo_concentrations} relation predicts for subhaloes nearer the host centre (Section~\ref{subsec:lens_subhalo}). More concentrated haloes, including those above the relation when scatter is included, could be detected at lower masses \citep{amorisco2022_halo_concentration}. The thresholds also describe sensitivity to individual subhaloes in the lens plane, and constraining dark matter would additionally require the abundance of perturbers, including haloes along the line of sight, and the selection of the observed lenses. Baryonic processes can also alter subhalo abundances and internal structure \citep{Porras2026_baryons}. Fits with reconstructed sources and lens light would test how much of the reach of the selected systems remains in a more complete analysis. Ranking simulated survey images would test whether the selection gain carries over to real candidates, and recovering PSFs from simulated HWO calibration observations would test whether the required modelling accuracy can be achieved. A search over the full arc with its own controls would similarly test the false-positive rate.

\section{Conclusions}
\label{sec:conclusion}

Following the HWO strong-lensing science case of \citet{he2026_hwo25_strong_lensing}, we have used HWO-SLAPS to forecast subhalo sensitivity for 965 simulated lens systems under a common 2000-s HWO EAC1/HRI observation, and tested how PSF quality and modelling errors affect the 12 highest-ranked systems. Ranking the systems by arc signal-to-noise and relative gradient power before calculating subhalo sensitivity gives a top-50 median $M_{\rm best}$ of $10^{7.51}\msun$, about 1.1~dex below the ensemble median, and $10^{7.19}\msun$ for the top 12. Selection also lowers the median $M_{10}$ from $10^{9.11}$ to $10^{8.12}\msun$, so the approximately one-dex gain extends to the mass detectable over 10\% of the aperture. Our ranking uses noise-free expected images, and realizing this gain in a follow-up programme depends on identifying the same arc signal-to-noise and relative gradient power in survey images.

Non-linear fits detect all 50 selected Asimov injections at their tested masses and positions. For the top 12, linear and non-linear calculations with the same image treatment closely agree, showing that finite lens and source adjustments change the detection statistic very little. The 12 validation systems that the Fisher maps predict as detectable but the non-linear fits miss are tested at $10^9$--$10^{9.5}\msun$, above every test mass in the selected sample. With one noise realization, 38 of the 50 selected injections are recovered, consistent with testing close to the detection threshold. These comparisons support using Fisher maps to explore many systems and observing configurations, with non-linear checks at selected masses and positions.

For the top 12 systems, increasing the physical wavefront error from 35 to 65~nm RMS raises the median mass thresholds by about 0.1~dex when the PSF is modelled correctly. Errors in the fitted PSF can instead remove detectable positions and increase the apparent preference for a subhalo, although only one of the 143 accepted PSF-error controls gives a false detection. Under our illustrative retention criteria, the median tolerated wavefront mismatch is 5~nm RMS, with a range of 2--10~nm across the systems. These forecasts assume known source structure and omit lens-galaxy light, so they are optimistic in these respects. The adopted subhalo concentrations are instead conservative, since subhaloes nearer the host centre or above the median relation would be easier to detect. These results support selective HWO follow-up of the large lens samples expected from Euclid and Roman. With accurate PSF modelling to preserve sensitivity to low-mass subhaloes, such a programme could transform our understanding of the nature of dark matter.

\section*{Acknowledgements}

GNV acknowledges support through the Habitable Worlds Observatory (HWO) internship at the Jet Propulsion Laboratory (JPL). SR and JR were supported by JPL, which is operated by the California Institute of Technology for NASA. RM was supported by the UK Space Agency through grants UKRI3163 and UKRI5725. QH was supported by the Dutch Research Council (NWO) through the grant Open Competitie ENW M21-1. Portions of this research were carried out at the Jet Propulsion Laboratory, California Institute of Technology, under a contract with the National Aeronautics and Space Administration (80NM0018D0004). The authors would like to thank Eric Huff and the Jet Propulsion Laboratory Section 3268 for helpful discussions and feedback on this work. 

ChatGPT and Codex (OpenAI) and Claude (Anthropic) were used to assist with code development, analysis, and manuscript outlining and editing. The authors take full responsibility for the content and conclusions of this work.

\section*{Data Availability}

HWO-SLAPS is publicly available at \url{https://github.com/nasa-jpl/hwo-slaps}. The study-specific configurations, analysis scripts, and simulation outputs are available from the corresponding author on reasonable request.

\section*{Conflict of Interest}

The authors declare no conflict of interest.

\bibliographystyle{rasti}
\bibliography{report}

\begin{thebibliography}{39}
\expandafter\ifx\csname natexlab\endcsname\relax\def\natexlab#1{#1}\fi

\bibitem[Amorisco et~al.(2022)Amorisco, Nightingale, He, Amvrosiadis, Cao,
  Cole, Etherington, Frenk, Li, Massey, \&
  Robertson]{amorisco2022_halo_concentration}
Amorisco, N.~C., Nightingale, J., He, Q., Amvrosiadis, A., Cao, X., Cole, S.,
  Etherington, A., Frenk, C.~S., Li, R., Massey, R., \& Robertson, A., 2022.
\newblock Halo concentration strengthens dark matter constraints in
  galaxy--galaxy strong lensing analyses, {\it Monthly Notices of the Royal
  Astronomical Society\/}, {\bf 510}(2), 2464--2479.

\bibitem[Bartelmann(2010)]{bartelmann2010_lensing}
Bartelmann, M., 2010.
\newblock Gravitational lensing, {\it Classical and Quantum Gravity\/}, {\bf
  27}(23), 233001.

\bibitem[Bode et~al.(2001)Bode, Ostriker, \& Turok]{bode2001_wdm}
Bode, P., Ostriker, J.~P., \& Turok, N., 2001.
\newblock Halo formation in warm dark matter models, {\it The Astrophysical
  Journal\/}, {\bf 556}, 93--107.

\bibitem[Bullock \& Boylan-Kolchin(2017)]{bullock2017_small_scale_challenges}
Bullock, J.~S. \& Boylan-Kolchin, M., 2017.
\newblock Small-scale challenges to the {$\Lambda$CDM} paradigm, {\it Annual
  Review of Astronomy and Astrophysics\/}, {\bf 55}, 343--387.

\bibitem[Byrd et~al.(1995)Byrd, Lu, Nocedal, \& Zhu]{byrd1995_lbfgsb}
Byrd, R.~H., Lu, P., Nocedal, J., \& Zhu, C., 1995.
\newblock A limited memory algorithm for bound constrained optimization, {\it
  SIAM Journal on Scientific Computing\/}, {\bf 16}(5), 1190--1208.

\bibitem[Cohen \& Kaplan(2007)]{cohen2007_bottom_k}
Cohen, E. \& Kaplan, H., 2007.
\newblock Summarizing data using bottom-{$k$} sketches, in {\em Proceedings of
  the 26th Annual ACM Symposium on Principles of Distributed Computing\/}, pp.
  225--234, Association for Computing Machinery.

\bibitem[Collett(2015)]{collett2015_euclid_predict}
Collett, T.~E., 2015.
\newblock The population of galaxy–galaxy strong lenses in forthcoming
  optical imaging surveys, {\it The Astrophysical Journal\/}, {\bf 811}(1), 20.

\bibitem[Cowan et~al.(2011)Cowan, Cranmer, Gross, \&
  Vitells]{cowan2011_asymptotic}
Cowan, G., Cranmer, K., Gross, E., \& Vitells, O., 2011.
\newblock Asymptotic formulae for likelihood-based tests of new physics, {\it
  The European Physical Journal C\/}, {\bf 71}(2), 1554.

\bibitem[Despali et~al.(2022)Despali, Vegetti, White, Powell, Stacey,
  Fassnacht, Rizzo, \& Enzi]{despali2022_sensitivity}
Despali, G., Vegetti, S., White, S. D.~M., Powell, D.~M., Stacey, H.~R.,
  Fassnacht, C.~D., Rizzo, F., \& Enzi, W., 2022.
\newblock Detecting low-mass haloes with strong gravitational lensing {I}: the
  effect of data quality and lensing configuration, {\it Monthly Notices of the
  Royal Astronomical Society\/}, {\bf 510}(2), 2480--2494.

\bibitem[Dressing et~al.(2026)]{dressing2026_hwo_discovery}
Dressing, C.~D. et~al., 2026.
\newblock Early exploration of the scientific discovery space for the
  {Habitable Worlds Observatory}, arXiv:2608.11294.

\bibitem[{Euclid Collaboration} et~al.(2026){Euclid Collaboration}, Walmsley,
  et~al.]{walmsley2026_euclid_dr0}
{Euclid Collaboration}, Walmsley, M., et~al., 2026.
\newblock {Euclid} quick data release ({Q1}) -- {XXVI}. the strong lensing
  discovery engine {A} -- system overview and first lens sample, {\it A\&A\/},
  {\bf 711}, A26.

\bibitem[He et~al.(2023)]{he2022_cosmo_subhalo_detection}
He, Q. et~al., 2023.
\newblock Testing strong lensing subhalo detection with a cosmological
  simulation, {\it Monthly Notices of the Royal Astronomical Society\/}, {\bf
  518}(1), 220--239.

\bibitem[He et~al.(2026)He, Lagattuta, Jauzac, Daylan, Massey, Rhodes, Doppel,
  Moustakas, Kiessling, Smith, Werner, Nightingale, von Wietersheim-Kramsta, \&
  Das]{he2026_hwo25_strong_lensing}
He, Q., Lagattuta, D., Jauzac, M., Daylan, T., Massey, R., Rhodes, J., Doppel,
  J.~E., Moustakas, L., Kiessling, A., Smith, R., Werner, S., Nightingale, J.,
  von Wietersheim-Kramsta, M., \& Das, S., 2026.
\newblock {Habitable Worlds Observatory}: Strong lensing constraints on dark
  matter halo mass function down to {$10^{7}\,M_{\odot}$}, in {\em {HWO25
  Proceedings Part I}: Community Science Case Development Documents\/}, vol.~10
  of {\bf ASP Conference Series Monograph}, p. 221, Astronomical Society of the
  Pacific.

\bibitem[Koekemoer et~al.(2007)]{koekemoer2007_cosmos_acs}
Koekemoer, A.~M. et~al., 2007.
\newblock The {COSMOS} survey: {Hubble Space Telescope}/{Advanced Camera for
  Surveys} ({HST}/{ACS}) observations and data processing, {\it The
  Astrophysical Journal Supplement Series\/}, {\bf 172}, 196.

\bibitem[Kormann et~al.(1994)Kormann, Schneider, \&
  Bartelmann]{kormann1994_sie}
Kormann, R., Schneider, P., \& Bartelmann, M., 1994.
\newblock Isothermal elliptical gravitational lens models, {\it Astronomy \&
  Astrophysics\/}, {\bf 284}, 285--299.

\bibitem[Lajoie et~al.(2023)Lajoie, Lallo, Mel{\'e}ndez, Flagey, Telfer,
  Comeau, Kulp, Beck, Brady, \& Perrin]{lajoie2023_jwst_wavefront}
Lajoie, C.-P., Lallo, M., Mel{\'e}ndez, M., Flagey, N., Telfer, R., Comeau,
  T.~M., Kulp, B.~A., Beck, T., Brady, G.~R., \& Perrin, M.~D., 2023.
\newblock A year of wavefront sensing with {JWST} in flight: Cycle 1 telescope
  monitoring and maintenance summary, OTE Science Performance Memo
  JWST-STScI-008497, Space Telescope Science Institute.

\bibitem[Lange(2023)]{lange2023_nautilus}
Lange, J.~U., 2023.
\newblock {NAUTILUS}: boosting {Bayesian} importance nested sampling with deep
  learning, {\it Monthly Notices of the Royal Astronomical Society\/}, {\bf
  525}(2), 3181--3194.

\bibitem[Liu et~al.(2026)]{liu2026_hwo_eacs}
Liu, A. K.-C. et~al., 2026.
\newblock Early architecture concepts for the {Habitable Worlds Observatory}
  system design, modeling, and analysis, {\it Journal of Astronomical
  Telescopes, Instruments, and Systems\/}, {\bf 12}(04), 041017.

\bibitem[Lovell et~al.(2014)Lovell, Frenk, Eke, Jenkins, Gao, \&
  Theuns]{lovell2014_wdm_haloes}
Lovell, M.~R., Frenk, C.~S., Eke, V.~R., Jenkins, A., Gao, L., \& Theuns, T.,
  2014.
\newblock The properties of warm dark matter haloes, {\it Monthly Notices of
  the Royal Astronomical Society\/}, {\bf 439}, 300--317.

\bibitem[Mahajan \& Dai(2006)]{mahajan2006_hexikes}
Mahajan, V.~N. \& Dai, G.-m., 2006.
\newblock Orthonormal polynomials for hexagonal pupils, {\it Optics Letters\/},
  {\bf 31}(16), 2462--2464.

\bibitem[Molin{\'e} et~al.(2017)Molin{\'e}, S{\'a}nchez-Conde, Palomares-Ruiz,
  \& Prada]{moline2017_subhalo_concentrations}
Molin{\'e}, {\'A}., S{\'a}nchez-Conde, M.~A., Palomares-Ruiz, S., \& Prada, F.,
  2017.
\newblock Characterization of subhalo structural properties and implications
  for dark matter annihilation signals, {\it Monthly Notices of the Royal
  Astronomical Society\/}, {\bf 466}(4), 4974--4990.

\bibitem[{National Academies of Sciences, Engineering, and
  Medicine}(2023)]{nasem2023_astro2020}
{National Academies of Sciences, Engineering, and Medicine}, 2023.
\newblock {\it Pathways to Discovery in Astronomy and Astrophysics for the
  2020s\/}, The National Academies Press, Washington, DC.

\bibitem[Navarro et~al.(1997)Navarro, Frenk, \& White]{navarro1997_nfw}
Navarro, J.~F., Frenk, C.~S., \& White, S. D.~M., 1997.
\newblock A universal density profile from hierarchical clustering, {\it The
  Astrophysical Journal\/}, {\bf 490}(2), 493--508.

\bibitem[Newton et~al.(2011)Newton, Marshall, Treu, Auger, Gavazzi, Bolton,
  Koopmans, \& Moustakas]{newton2011_slacs}
Newton, E.~R., Marshall, P.~J., Treu, T., Auger, M.~W., Gavazzi, R., Bolton,
  A.~S., Koopmans, L. V.~E., \& Moustakas, L.~A., 2011.
\newblock The {Sloan} lens {ACS} survey. {XI}. beyond {Hubble} resolution:
  Size, luminosity, and stellar mass of compact lensed galaxies at intermediate
  redshift, {\it The Astrophysical Journal\/}, {\bf 734}(2), 104.

\bibitem[Nightingale et~al.(2021{\natexlab{a}})]{nightingale2021_pyautolens}
Nightingale, J.~W. et~al., 2021{\natexlab{a}}.
\newblock {PyAutoLens}: Open-source strong gravitational lensing, {\it Journal
  of Open Source Software\/}, {\bf 6}(58), 2825.

\bibitem[Nightingale et~al.(2024)]{nightingale2023_subhalos}
Nightingale, J.~W. et~al., 2024.
\newblock Scanning for dark matter subhaloes in {Hubble Space Telescope}
  imaging of 54 strong lenses, {\it Monthly Notices of the Royal Astronomical
  Society\/}, {\bf 527}(4), 10480--10506.

\bibitem[Nightingale et~al.(2021{\natexlab{b}})Nightingale, Hayes, \&
  Griffiths]{nightingale2021_pyautofit}
Nightingale, J.~W., Hayes, R.~G., \& Griffiths, M., 2021{\natexlab{b}}.
\newblock {PyAutoFit}: A classy probabilistic programming language for model
  composition and fitting, {\it Journal of Open Source Software\/}, {\bf
  6}(58), 2550.

\bibitem[O'Riordan et~al.(2023)O'Riordan, Despali, Vegetti, Lovell, \&
  Molin{\'e}]{oriordan2023_euclid_sensitivity}
O'Riordan, C.~M., Despali, G., Vegetti, S., Lovell, M.~R., \& Molin{\'e},
  {\'A}., 2023.
\newblock Sensitivity of strong lensing observations to dark matter
  substructure: a case study with {Euclid}, {\it Monthly Notices of the Royal
  Astronomical Society\/}, {\bf 521}(2), 2342--2356.

\bibitem[{Planck Collaboration} et~al.(2016){Planck Collaboration}, Ade,
  et~al.]{planck2015_cosmological_parameters}
{Planck Collaboration}, Ade, P. A.~R., et~al., 2016.
\newblock {Planck} 2015 results. {XIII}. cosmological parameters, {\it
  Astronomy \& Astrophysics\/}, {\bf 594}, A13.

\bibitem[Por et~al.(2018)Por, Haffert, Radhakrishnan, Doelman, Van~Kooten, \&
  Bos]{por2018_hcipy}
Por, E.~H., Haffert, S.~Y., Radhakrishnan, V.~M., Doelman, D.~S., Van~Kooten,
  M., \& Bos, S., 2018.
\newblock High contrast imaging for {Python} ({HCIPy}): an open-source adaptive
  optics and coronagraph simulator, in {\em Adaptive Optics Systems VI\/}, vol.
  10703 of {\bf Proc. SPIE}, p. 1070342.

\bibitem[Porras-Bedmar et~al.(2026)Porras-Bedmar, S{\'a}nchez-Conde, \&
  Aguirre-Santaella]{Porras2026_baryons}
Porras-Bedmar, S., S{\'a}nchez-Conde, M.~{\'A}., \& Aguirre-Santaella, A.,
  2026.
\newblock Impact of baryons on the population of galactic subhalos and
  implications for dark matter searches, {\it Physics of the Dark Universe\/},
  {\bf 52}, 102262.

\bibitem[Rowe et~al.(2015)]{rowe2015_galsim}
Rowe, B. T.~P. et~al., 2015.
\newblock {GalSim}: The modular galaxy image simulation toolkit, {\it Astronomy
  and Computing\/}, {\bf 10}, 121--150.

\bibitem[{The LUVOIR Team}(2019)]{luvoir2019_final_report}
{The LUVOIR Team}, 2019.
\newblock The {LUVOIR} mission concept study final report, arXiv:1912.06219.

\bibitem[Vassilakis et~al.(2026)Vassilakis, Redmond, Rhodes, Massey, \&
  He]{vassilakis2026_hwo_slaps}
Vassilakis, G.~N., Redmond, S.~F., Rhodes, J.~D., Massey, R.~J., \& He, Q.,
  2026.
\newblock Forecasting {PSF} stability requirements for low-mass dark matter
  subhalo detection with the {Habitable Worlds Observatory}, in {\em Space
  Telescopes and Instrumentation 2026: Optical, Infrared, and Millimeter
  Wave\/}, vol. 14145 of {\bf Proceedings of SPIE}, p. 141456R, SPIE.

\bibitem[Vegetti \& Koopmans(2009)]{vegetti2009_bayesian}
Vegetti, S. \& Koopmans, L. V.~E., 2009.
\newblock Bayesian strong gravitational-lens modelling on adaptive grids:
  objective detection of mass substructure in galaxies, {\it Monthly Notices of
  the Royal Astronomical Society\/}, {\bf 392}(3), 945--963.

\bibitem[Vegetti et~al.(2010)Vegetti, Koopmans, Bolton, Treu, \&
  Gavazzi]{vegetti2010_dark_substructure}
Vegetti, S., Koopmans, L. V.~E., Bolton, A., Treu, T., \& Gavazzi, R., 2010.
\newblock Detection of a dark substructure through gravitational imaging, {\it
  Monthly Notices of the Royal Astronomical Society\/}, {\bf 408}(4),
  1969--1981.

\bibitem[Vegetti et~al.(2014)Vegetti, Koopmans, Auger, Treu, \&
  Bolton]{vegetti2014_inference}
Vegetti, S., Koopmans, L. V.~E., Auger, M.~W., Treu, T., \& Bolton, A.~S.,
  2014.
\newblock Inference of the cold dark matter substructure mass function at z =
  0.2 using strong gravitational lenses, {\it Monthly Notices of the Royal
  Astronomical Society\/}, {\bf 442}(3), 2017--2035.

\bibitem[Virtanen et~al.(2020)Virtanen, Gommers, Oliphant,
  et~al.]{virtanen2020_scipy}
Virtanen, P., Gommers, R., Oliphant, T.~E., et~al., 2020.
\newblock {SciPy} 1.0: fundamental algorithms for scientific computing in
  {Python}, {\it Nature Methods\/}, {\bf 17}, 261--272.

\bibitem[Wedig et~al.(2025)Wedig, Daylan, Birrer,
  et~al.]{wedig2025_roman_lenses}
Wedig, B., Daylan, T., Birrer, S., et~al., 2025.
\newblock The {Roman} view of strong gravitational lenses, {\it The
  Astrophysical Journal\/}, {\bf 986}(1), 42.

\end{thebibliography}

\clearpage
\appendix
\section{Lens population sampling}
\label{app:population_sampling}

\begingroup
\makeatletter
\def\fps@table{htbp}
\makeatother
\begin{table}
\centering
\caption{Initial-population draws before observability cuts. $\mathcal{N}(\mu,\sigma^2)$ denotes a normal distribution; $\mathcal{L}(m,s)$ a log-normal with median $m$ and natural-log standard deviation $s$, both before truncation to the listed bounds. Here $z_{\min}=\max(0.35,z_{\rm l}+0.20)$, and $m_{\rm AB}$ is the unlensed HRI-V magnitude.}
\label{tab:population_parameters}
\small
\setlength{\tabcolsep}{3pt}
\begin{tabular}{@{}p{0.30\columnwidth}p{0.38\columnwidth}p{0.26\columnwidth}@{}}
\toprule
Parameter & Distribution & Bounds \\
\midrule
\multicolumn{3}{@{}l}{\textbf{Lens and redshift geometry}} \\
\addlinespace[0.4ex]
Lens $z_{\rm l}$ & $\mathcal{N}(0.22,0.09^2)$ & $0.08$--$0.50$ \\
Source $z_{\rm s}$ & $\mathcal{N}(0.65,0.22^2)$ & $z_{\min}$--$1.20$ \\
$\sigma_v$ ($\mathrm{km\,s^{-1}}$) & $\mathcal{L}(245,0.18)$ & $130$--$360$ \\
Axis ratio $q_{\rm lens}$ & $\mathcal{N}(0.75,0.12^2)$ & $0.40$--$0.95$ \\
Position angle & Uniform & $0$--$180^{\circ}$ \\
\addlinespace[0.6ex]
\multicolumn{3}{@{}l}{\textbf{Source placement and properties}} \\
\addlinespace[0.4ex]
Offset $\beta/\theta_{\rm E}$ & Log-uniform & $0.05$--$1.00$ \\
Centre azimuth & Uniform & $0$--$360^{\circ}$ \\
$m_{\rm AB}$ & $\mathcal{N}(24.845,0.80^2)$ & $23.745$--$27.545$ \\
Size $r_{\rm h}$ (arcsec) & $\mathcal{L}(0.15,0.35)$ & $0.06$--$0.40$ \\
Rotation & Uniform & $0$--$360^{\circ}$ \\
Template & \multicolumn{2}{l}{200 systems per template} \\
\bottomrule
\end{tabular}
\end{table}
\endgroup

We construct the validation sample using bottom-$k$ sampling \citep{cohen2007_bottom_k}. The 965 surviving systems are ordered by the SHA-256 hashes of their identifiers and the first $k=100$ are retained.

\begingroup
\makeatletter
\def\fps@table{htbp}
\makeatother
\begin{table}
\centering
\caption{Angular areas corresponding to 10\% aperture coverage for the full ensemble and samples selected by the pre-detection ranking $\mathcal{R}$. All areas are in $\mathrm{arcsec}^{2}$ and are calculated from the sampled aperture cells. The corresponding 50\% areas are five times larger. The full-ensemble summary includes systems whose $M_{10}$ threshold is not reached within the sampled mass range.}
\label{tab:aperture_areas}
\small
\setlength{\tabcolsep}{3pt}
\begin{tabular*}{\columnwidth}{@{\extracolsep{\fill}}lrrrr@{}}
\toprule
Sample & $N$ & Median & Mean & Range \\
\midrule
Full ensemble & 965 & 1.50 & 1.97 & 0.32--11.72 \\
Top 50 & 50 & 1.37 & 1.90 & 0.44--5.48 \\
Top 12 & 12 & 1.37 & 2.02 & 0.60--4.65 \\
\bottomrule
\end{tabular*}
\par\smallskip
\parbox{\columnwidth}{\footnotesize Medians are the smallest values at which the empirical cumulative fraction reaches 0.5.}
\end{table}
\endgroup

\section{Numerical procedure of the non-linear fits}
\label{app:nonlinear_numerics}

The Nautilus searches in Section~\ref{subsec:nonlinear_comparisons} use 300 live points for the smooth model and 600 for the subhalo model, with a minimum effective-sample target of $n_{\rm eff}=2000$ and a limit of $2\times10^6$ likelihood evaluations per model. For optimization, we rescale each parameter to $[0,1]$ using its prior bounds (Table~\ref{tab:nonlinear_priors}). The sampler's maximum-likelihood point supplies one starting point. We select eight further samples in descending likelihood order, requiring separation from every previously selected point by a Euclidean distance of at least 0.05 in these coordinates, or at least one after scaling each coordinate by its weighted posterior standard deviation.

From each starting point, bounded L-BFGS-B minimizes $\chi^2/2$ using gradients obtained by automatic differentiation. Each run allows 500 iterations and 50 line-search steps per iteration, with a gradient tolerance of $10^{-10}$ and zero function tolerance. We retain the highest likelihood encountered across the runs, then repeat the optimization from that point with 1000 iterations and a gradient tolerance of $10^{-12}$. We accept a maximum only if at least two of the eight additional starts reach log likelihoods within 0.1 of it and the tighter repeat changes it by no more than 0.1. The retained point must have a finite gradient. Its noise-weighted residual sum of squares must agree with the optimizer's $\chi^2$ to within $10^{-4}$, and its directly evaluated log likelihood must agree with that implied by the optimizer to the same tolerance. We also check that the sampler's maximum reproduces its saved log likelihood to within $10^{-4}$.

If either model fails these checks, we repeat both searches once with a fresh sampler seed, doubling the number of live points, the effective-sample target, and the likelihood-evaluation limit. This gives 600 live points for the smooth model, 1200 for the subhalo model, $n_{\rm eff}=4000$, and a limit of $4\times10^6$ likelihood evaluations per model. The images, priors, refinement procedure, and acceptance checks remain the same. We adopt this second comparison only if both models pass and neither log-likelihood maximum falls more than 0.1 below its first-pass value. Otherwise, the comparison remains unresolved and is excluded from the detection counts.

For each comparison, we retain the input configurations, random seeds, code revision, sampled parameter vectors, optimized likelihoods, and numerical checks. The records also identify the source template and PSF kernels by their hashes, linking each reported statistic to its inputs and fitting procedure.

\bsp
\label{lastpage}
\end{document}